\documentstyle[preprint,aps,epsf]{revtex}
\tighten
\newcommand{\bold}[1]{\mbox{\boldmath $#1$}}    
\newcommand{\ul}[1]{\underline{#1}}             
\newcommand{\del}{\partial}                     
\newcommand{\PACS}[1]{~\\ \noindent{#1}}        
\newcommand{\bolleaa}{\raisebox{1.2ex}         
        {\hspace{0.2ex}\scriptsize{$\circ$}}\hspace{-1.1ex}{a}}
\newcommand{\etal}{{\em et al.}}                
\newcommand{\ie}{{\em i.e.}}                    
\newcommand{\MeV}{{\rm MeV}}                    
\newcommand{\fm}{{\rm fm}}                      
\newcommand{\rme}{{\rm e}}                      
\newcommand{\k}{{\bf{k}}}                       
\newcommand{\r}{{\bold{r}}}                     
\newcommand{\p}{{\bold{p}}}                     
\newcommand{\half}{\mbox{${1\over2}$}}          
\newcommand{\third}{\mbox{${1\over3}$}}         
\newcommand{\lambdaq}{\mbox{${\lambda\over4}$}} 
\newcommand{\cchi}{{\bold{\chi}}}               
\newcommand{\pphi}{{\bold{\phi}}}               
\newcommand{\ppsi}{{\bold{\psi}}}               
\newcommand{\dpphi}{{\bold{\delta\phi}}}        
\newcommand{\dppsi}{{\bold{\delta\psi}}}        
\newcommand{\dphi}{\delta\phi}                  
\newcommand{\x}{{\mbox{\boldmath{{$\rho$}}}}}
\newcommand{\xscript}{{\mbox{\boldmath{\scriptsize{$\rho$}}}}}

\begin{document}
\begin{titlepage}
\noindent{\sl Physical Review C}\hfill LBNL-43549\\[8ex]

\begin{center}
{\large {\bf Dynamical simulation of DCC formation in Bjorken rods$^*$}}\\[8ex]
{\sl Troels C. Petersen and J\o rgen Randrup}\\[1ex]

Nuclear Science Division, Lawrence Berkeley National Laboratory\\
University of California, Berkeley, California 94720
\\[6ex]
July 9, 1999\\[6ex]
{\sl Abstract:}\\
\end{center}

{\small\noindent
Using a semi-classical treatment of the linear $\sigma$ model,
we simulate the dynamical evolution of an initially hot cylindrical rod
endowed with a longitudinal Bjorken scaling expansion (a ``Bjorken rod'').
The field equation is propagated until full decoupling has occurred
and the asymptotic many-body state of free pions
is then obtained by a suitable Fourier decomposition of the field
and a subsequent stochastic determination of the number of quanta
in each elementary mode.
The resulting transverse pion spectrum exhibits visible enhancements
below 200~MeV due to the parametric amplification
caused by the oscillatory relaxation of the chiral order parameter.
Ensembles of such final states 
are subjected to various event-by-event analyses.
The factorial moments of the multiplicity distribution
suggest that the soft pions are non-statistical.
Furthermore,
their emission patterns exhibit azimuthal correlations
that have a bearing on the domain size in the source.
Finally,
the distribution of the neutral pion fraction shows a significant broadening
for the soft pions
which grows steadily as the number of azimuthal segments is increased.
All of these features are indicative of disoriented chiral condensates
and it may be interesting to apply similar analyses
to actual data from high-energy nuclear collision experiments.
\\[4ex]


\vfill
\noindent
$^*$This work was supported by the Director, Office of Energy Research,
Office of High Energy and Nuclear Physics,
Nuclear Physics Division of the U.S. Department of Energy
under Contract No.\ DE-AC03-76SF00098.
TCP was supported in part by the Danish Ministry of Education,
the Danish Ministry of Research,
H{\o}jg\bolleaa rd Fonden, and L{\o}rups Mindelegat.
}
\end{titlepage}


\section{Introduction}

The possibility of forming {\em disoriented chiral condensates} (DCC)
in high-energy nuclear collisions has gained significant attention
in recent years as a means for testing our understanding
of chiral symmetry restoration \cite{Rajagopal:QGP2,BK:review,bj97}.
A crossover from the normal phase,
in which chiral symmetry is broken,
to a phase with approximately restored chiral symmetry
is expected to occur at temperatures of a few hundred MeV.
Such energy densities are readily reached during the early stage
of an ultrarelativistic nuclear collision
and one may therefore expect that approximate restoration
of chiral symmetry occurs transiently in the hot collision zone.
The subsequent relaxation towards the normal vacuum
has a non-equilibrium character
due to the very rapid expansion of the system.
As a result,
the pion field acquires large-amplitude long-wavelength oscillations
which may have observable consequences.
The suggested signals include an excess of isospin-aligned soft pions
with an associated anomalously broad distribution of the neutral pion fraction
\cite{Anselm:PLB217,Anselm:PLB266,Alaska,DCC,Wilczek:NPB399,Blaizot:PRD46}
and, in the electromagnetic sector,
a significant excess of photons \cite{Boyanovsky:PRD56b}
and dileptons \cite{zhxw96,KKRW}.

A few experimental DCC searches have already been carried out
\cite{MiniMax,WA98}
but no signals were discerned so far.
On the theoretical side,
a considerable number of explorative studies of this hypothetical phenomenon
have been made over the past several years.
However, due to the complicated nature of the problem,
most studies have treated fairly idealized scenarios.
Therfore the practical observability of the phenomenon is not yet clarified.
The purpose of the present work is to treat a somewhat more refined scenario
and thus obtain a better basis for making an assessment of the prospects
for observing DCC signals and, in the process,
provide some guidance for the analysis of the experimental data.

The most popular tool for DCC studies has been the
SU(2) linear $\sigma$ model
which describes the $O(4)$ chiral field $\pphi=(\sigma,\bold{\pi})$
by means of a simple effective quartic interaction.
This framework has provided instructive insight into the features
of the non-equilibrium DCC dynamics
\cite{Rajagopal:NPB404,Blaizot:PRD50,GGP93,GM,Huang:PRD49,%
Cooper:PRD51,AHW:PRL74,Boyanovsky:PRD51,Csernai,MM,%
JR:PRD,Cooper:PRC54,Lampert,JR:PRL,JR:NPA,Kapusta:ZPC75,%
Biro:PRD55,GAC:PRD56,Boyanovsky:PRD56a}.
The present study is based on a semi-classical treatment
of the standard SU(2) linear $\sigma$ model \cite{JR:PRD}.
The dynamical variable is thus the $O(4)$ chiral field
$\pphi(\r)=(\sigma,\bold{\pi})$
which is subject to a self-interaction of the form
$V=\lambdaq[\phi^2-v^2]^2-H\sigma$ and, consequently,
it is governed by a second-order non-linear equation of motion,
\begin{equation}\label{EoM}
[\Box\ +\ \lambda(\phi^2-v^2)]{\pphi}\ =\ H\bold{e}_\sigma\ ,
\end{equation}
where $\bold{e}_\sigma$ denotes a unit vector in the $\sigma$ direction.
(Here and throughout we often omit factors of $\hbar$ and $c$
in order not to clutter the formulas.)

In the present study,
we wish to consider idealized scenarios that exhibits
some of the most important features expected in real collision events,
namely rapid longitudinal expansion and finite transverse extension.
Generally,
the numerical propagation of fields exhibiting significant flow patterns,
such as rapid expansion,
is practically difficult
due to the phase oscillations caused by the local boost.
However, for idealized scaling expansions
this complication can be eliminated by suitable variable transformations.
For the longitudinal scaling expansion considered here \cite{Bjorken},
it is convenient to replace the usual fixed-frame space-time variables
$(x,y,z,t)$
with the comoving variables $(x,y,\eta,\tau)$,
\begin{equation}\label{tz}
t=\tau\cosh\eta\ ,\ z=\tau\sinh\eta\ .
\end{equation}
Thus,
$\tau=(t^2-z^2)^{1/2}$ is the proper time experienced in a system
boosted along the $z$ axis with the local rapidity $\sf y$
equal to the value of $\eta=\half\ln({t+z \over t-z})$.
The corresponding form of the field equation of motion
is then obtained from the transformation of the d'Alembert operator,
\begin{equation}
\Box\ \equiv\ \partial_t^2-\partial_x^2-\partial_y^2-\partial_z^2\
=\ {1\over\tau}\del_\tau \tau\del_\tau
-\partial_x^2-\partial_y^2-{1\over\tau^2}{\del_\eta^2}\ .
\end{equation}
The equation of motion can be readily solved numerically
by application of the leapfrog method,
once the initial field $\pphi(\r)$
and its time derivative $\ppsi(\r)$ are specified.
(To help avoid confusion between the transverse coordinate $y$
and the rapidity $\sf y$,
we shall preferentially denote the position in the transverse plane
by $\x=(x,y)$ in the following,
in analogy with the use of $\k=(k_x,k_y)$ for the transverse wave number.)

\section{Bjorken matter}

We wish to first consider a macroscopically uniform system,
corresponding to infinite matter 
undergoing a longitudinal Bjorken scaling expansion.
Without the scaling expansion,
such a system would model infinite uniform matter.
We refer to the present expanding generalization as {\em Bjorken matter}
to capture the essential character of the system by a simple term.
Just as ordinary matter approximates conditions prevailing
in the bulk region of a finite system, such as a large atomic nucleus,
Bjorken matter approximates the conditions
in the interior of the rapidly stretching firestreak generated
in a central high-energy collision of two large nuclei.

For the numerical calculations,
we represent the field on a cartesian lattice in space
and impose periodic boundary conditions in all three directions.
The initial field configuration is prepared by means of the sampling
method developed in Ref.~\cite{JR:PRD}.
Thus an initial temperature $T_0$ is specified
and a field $\pphi'(\r)$ and its time derivative $\ppsi'(\r)$
are sampled from the corresponding thermal ensemble.
This field configuration represents a non-expanding system
in thermal equilibrium.
In order to obtain an initial field configuration 
suitable for the expansion scenario,
the coordinate $z$ is subsequently identified with 
the longitudinal variable $\eta$, 
\begin{eqnarray}
\pphi(\x,\eta,\tau_0)\ &\equiv&\ \pphi'(\x,z=\eta)\ ,\\
\ppsi(\x,\eta,\tau_0)\ &\equiv&\ \ppsi'(\x,z=\eta)\ ,
\end{eqnarray}
where we use $\tau_0=1~\fm/c$ for the initial value of the proper time,
in accordance with common practice.
In this manner it is assured that the local environment at $(\x,\eta)$,
when analyzed in a frame boosted with a rapidity $\sf y$ equal to $\eta$,
approximates that of matter in equilibrium at the specified temperature $T_0$.

\subsection{Dynamical evolution}

\begin{figure}
~\vspace{180mm}
\includegraphics{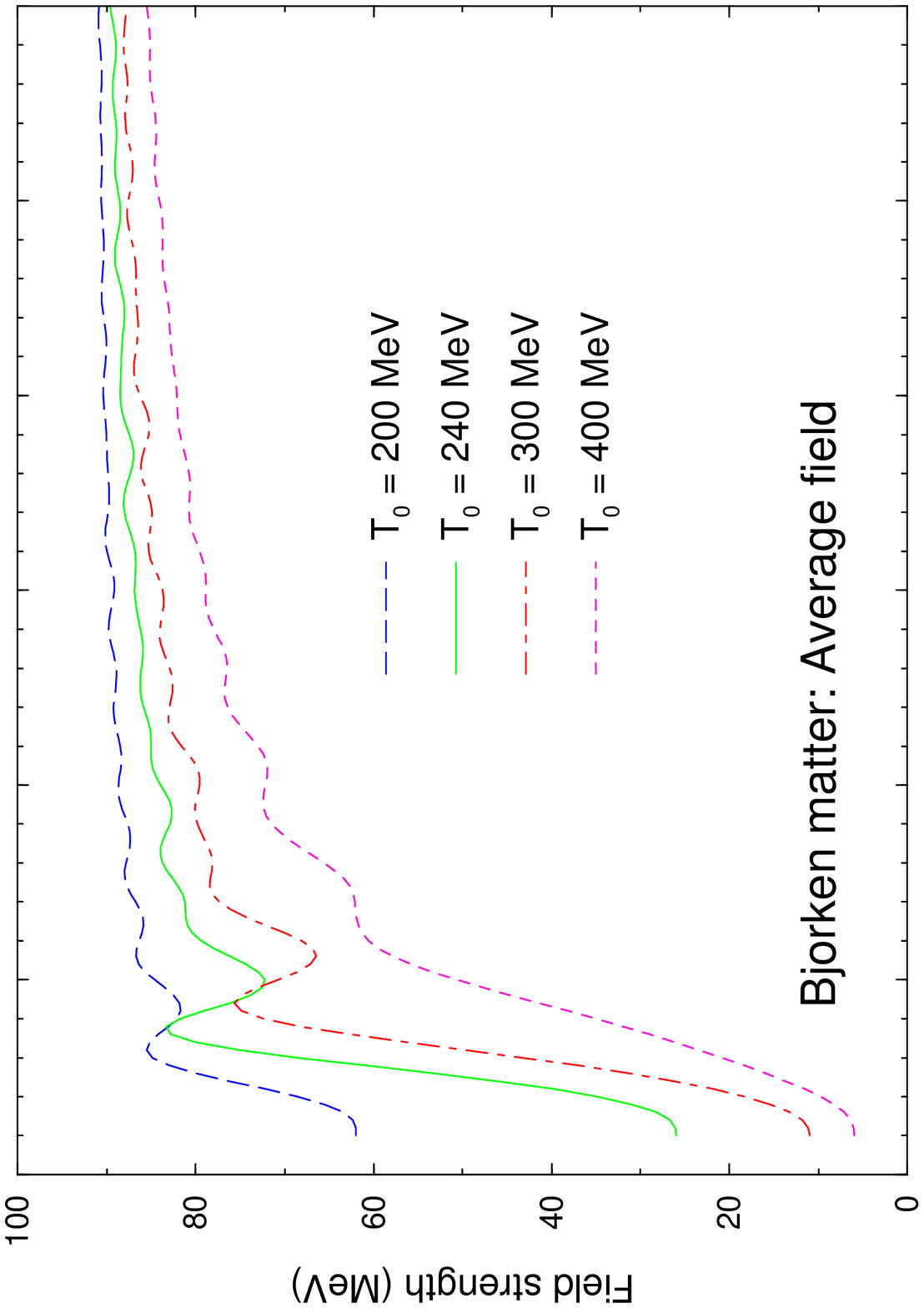}
\includegraphics{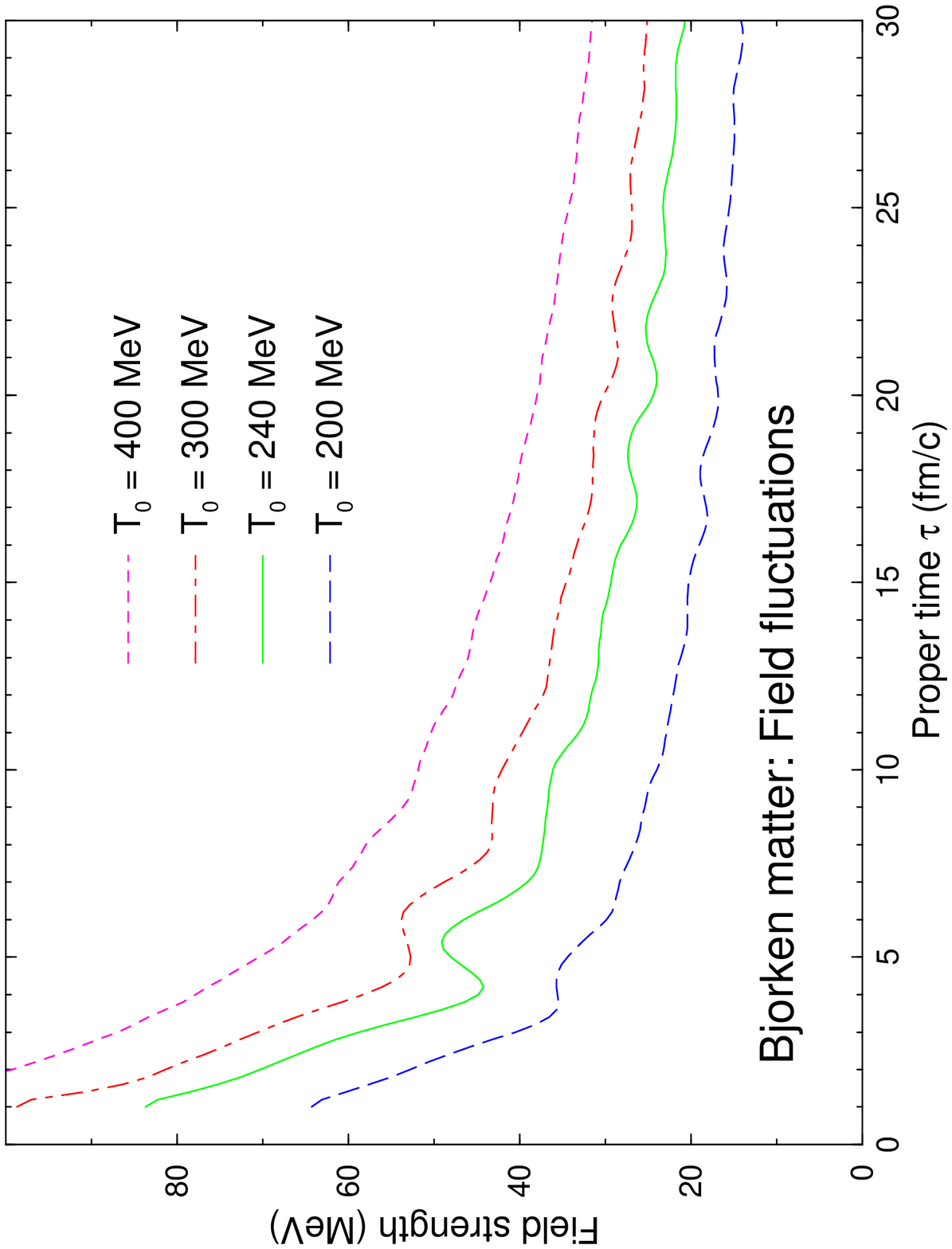}
\caption{Time evolution of the field.}
\label{f:time}
{\small
The time dependence of $\phi_0$,
the magnitude of the order parameter $\phi_0$ (top panel),
and $\delta\phi$,
the associated dispersion of the field fluctuations (bottom panel),
in Bjorken matter (a longitudinally expanding box)
prepared at various initial temperatures, $T_0$, as indicated.
}
\end{figure}

It is instructive to note the evolution of some of the key quantities.
Most important is the time dependence of the average field strength,
$\ul{\pphi}\equiv\langle\pphi\rangle$,
and the associated dispersion of the field fluctuations around that average,
$\delta\phi\equiv[\langle\phi^2\rangle-\phi_0^2]^{1/2}$,
where $\phi_0\equiv|\ul{\pphi}|$.
The former represents the O(4) chiral {\em order parameter},
while the latter reflects the presence of quasiparticle excitations
relative to that constant background field.
The dynamical evolution of these quantities is displayed in Fig.~\ref{f:time}
for a range of initial temperatures.
Initially, at $\tau=\tau_0$,
the system is hot and the order parameter is significantly reduced
relative to its vacuum value of $f_\pi=92~\MeV$,
with the field fluctuations being substantial
(and the more so the higher the value of $T_0$).
The longitudinal expansion effectively cools the system,
causing the field fluctuations to subside
(the field dispersion $\delta\phi$ evolves approximately
as $1/\sqrt{\tau}$ at large times).
This in turn makes the effective potential gradually revert to its vacuum form
and the order parameter exhibits a corresponding relaxation.
The specific oscillatory evolution is a result of the delicate balance
between the expansion rate
and oscillation frequency of the evolving effective potential.
In the intermediate temperature range where the phase cross over occurs
the order parameter exhibits a significant overshoot.
These oscillations in the order parameter are reflected in
the effective mass tensor for the quasiparticle excitations \cite{JR:NPA}
and, accordingly, the field fluctuations
(which depend inversely on the effective mass) exhibit corresponding
undulations superimposed on the overall steady decrease in time.

\begin{figure}
~\vspace{180mm}
\includegraphics{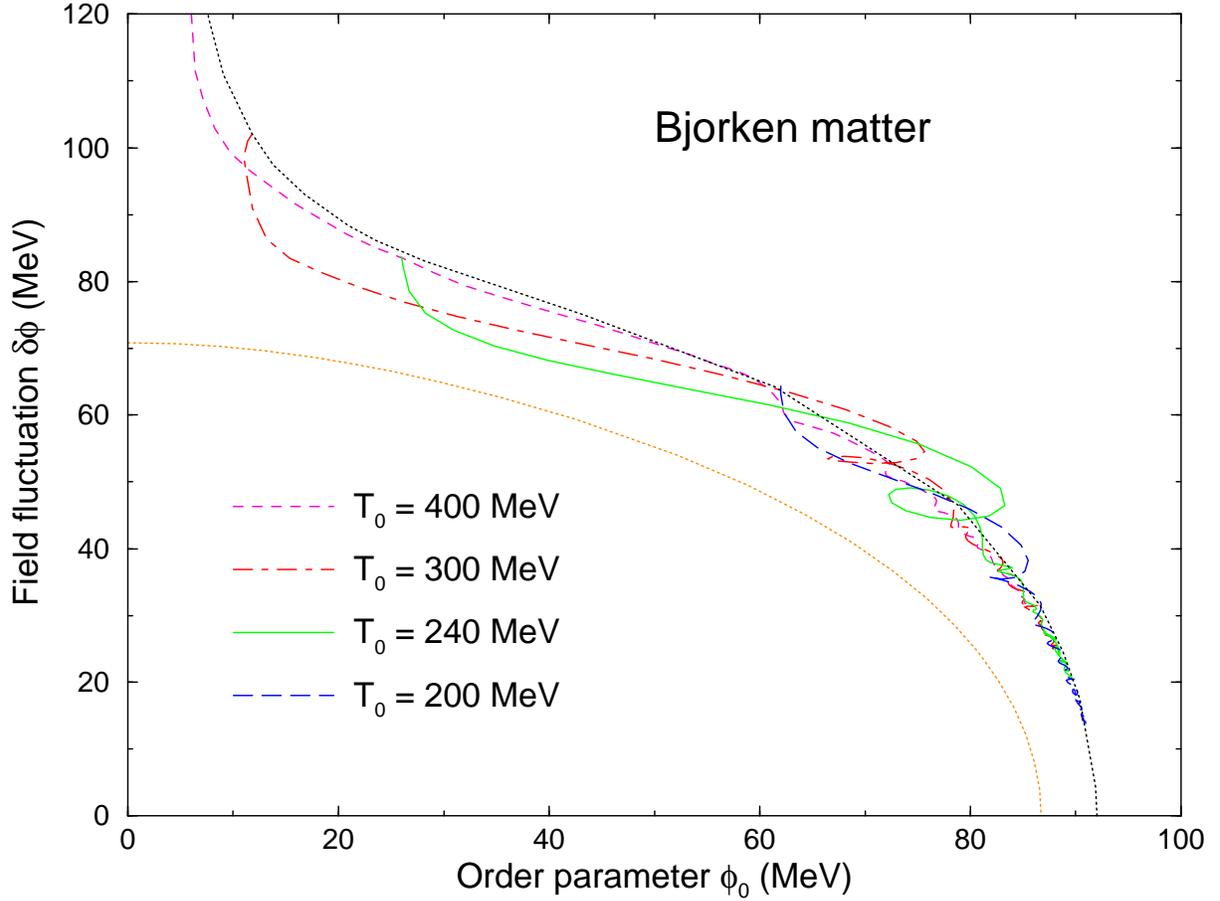}
\caption{Phase evolution.}
\label{f:phaseBox}
{\small
The time evolution of the expanding box
is projected onto the chiral phase diagram,
in which the abscissa is the order parameter $\phi_0$
(the magnitude of the average value of the chiral field within the box)
and the ordinate is the filed fluctuation $\delta\phi$
(the dispersion in the field around its average value),
for the four different initial temperatures $T_0$.
The equilibrium path is shown by the upper dotted curve
while the lower dotted curve delineates the region of instability
within which the field is supercritical.
}
\end{figure}

It is helpful to present the combined information in Fig.~\ref{f:time}
as a projection onto the chiral phase diagram,
in which the order parameter is the abscissa 
and the field fluctuation is the ordinate.
The former characterizes the medium-modified vacuum
and the latter is a measure of the degree of excitation
relative to that vacuum.
The result is displayed in Fig.~\ref{f:phaseBox}.
This figure includes the curve connecting the equilibrium points,
starting from ${\pphi}_{\rm gs}=(f_\pi,\bold{0})$ at $T=0$
and approaching the vertical axis at large temperatures
as chiral symmetry is gradually approached.
This curve represents the adiabatic path
along which the system would evolve if the expansion were sufficiently slow.
We note that the scaling expansion quickly brings the system away from
equilibrium, as the cooling of the field fluctuations is immediate
while the order parameter requires some time to adjust
to the changed effective potential.
This inertia leaves the order parameter on the inside of the minimum
in the effective potential and it therefore overshoots towards the outside
once it gets moving.
This pattern is repeated subsequently but on a smaller scale.
As a result, the evolution is fairly soon confined to within
a narrow neighborhood around the adiabatic path.
Nevertheless, the oscillations in the order parameter persist
for a long time with a slowly decreasing amplitude.
Included on the display is also the critical boundary within which
the field is supercritical and spontaneous pair creation occurs \cite{JR:PRD}.
We note that the dynamical paths stay safely away from this boundary,
regardless of the value of $T_0$,
so that no signals of supercriticality should be expected
for purely longitudinal expansions,
as has already been noted in earlier studies \cite{AHW:PRL74,JR:PRL}.

\begin{figure}
~\vspace{180mm}
\includegraphics{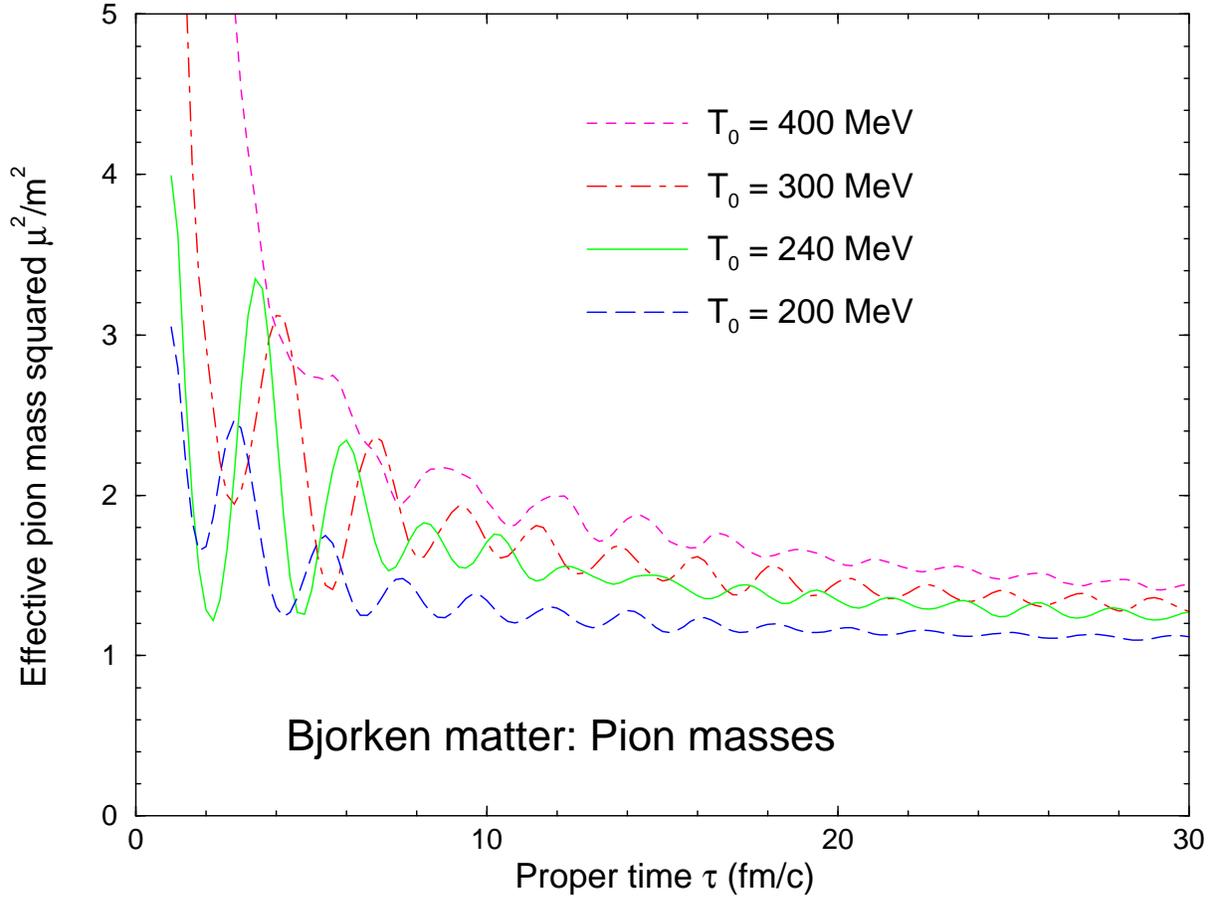}
\caption{Effective mass in Bjorken matter.}
\label{f:mass}
{\small
The time evolution of the square of the effective pion mass, $\mu_\pi^2(t)$
(divided by the square of its free mass $m_\pi$)
for Bjorken matter prepared at four different initial temperatures $T_0$
at the proper time $\tau_0=1~\fm/c$.
}
\end{figure}

In order to get an idea of what outcome might be expected,
it is helpful to consider the time evolution of the effective pion mass,
\begin{equation}\label{mu}
\mu_\pi^2(t)\ =\ \lambda[\phi_0^2+\delta\phi^2+2\delta\pi^2-v^2]\ ,
\end{equation}
where $\delta\phi^2=\delta\sigma^2+\delta\pi_1^2+\delta\pi_2^2+\delta\pi_3^2$
is the total field fluctuation and
$\delta\pi^2$ denotes the contribution from the field component
in the particular isospin direction of the pion considered.
This quantity is easily extracted as a by-product of the dynamical calculation
and it is displayed in Fig.~\ref{f:mass}.
Since the phase trajectory stays away from the critical region
(see Fig.~\ref{f:phaseBox}),
$\mu_\pi^2$ remains positive throughout.
Its evolution can be characterized as an overall steady decrease
(approximately inversely proportional to $\tau$)
superimposed by a fairly regular oscillation.

\begin{figure}
~\vspace{180mm}
\includegraphics{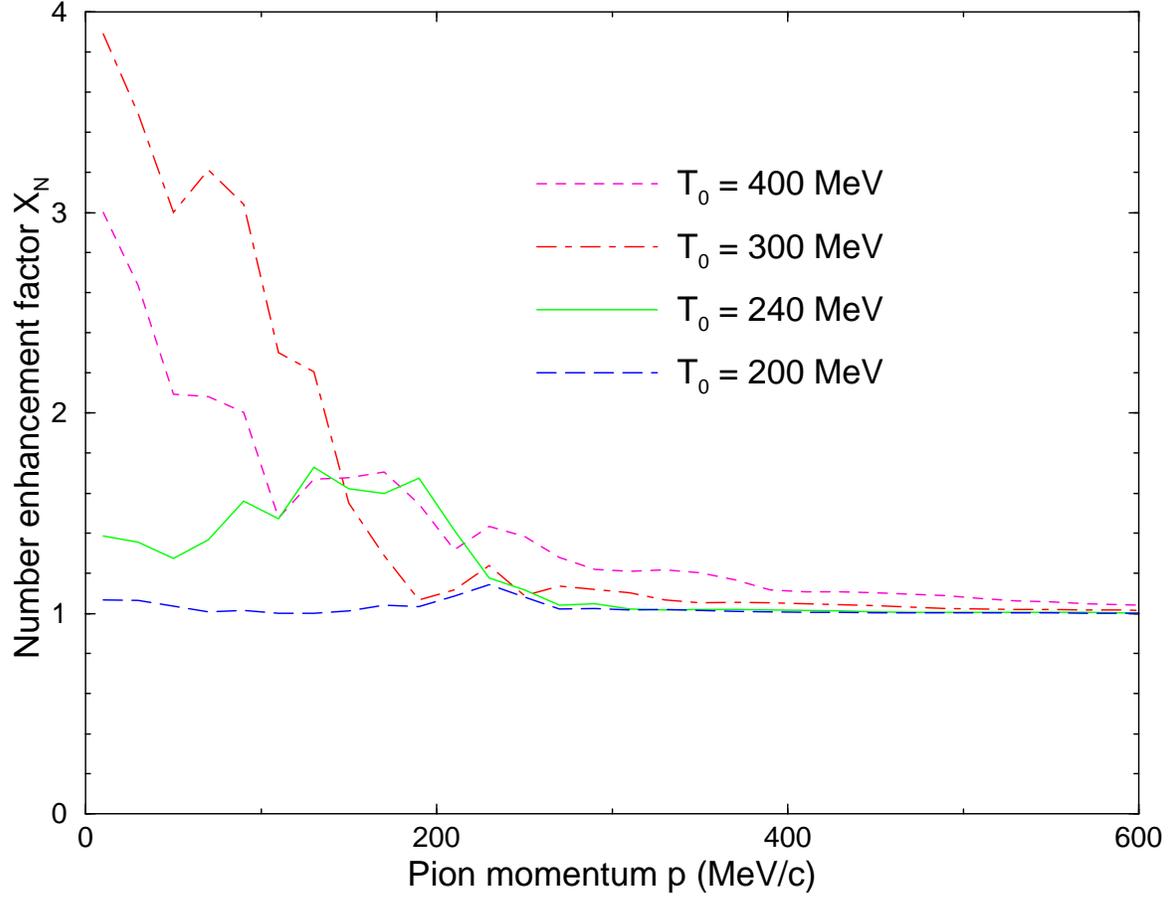}
\caption{Enhancement factor.}
\label{f:X}
{\small
The number enhancement coefficient $X_N$
obtained as described in Ref.~\cite{JR:HIP}
on the basis of the time dependent effective masses given in Fig.~\ref{f:mass}.
The coefficient $X_N$ expresses the relative increase in the population
of a given pion mode, as characterized by the momentum $p=\hbar kc$.
}
\end{figure}

It is often instructive to discuss the physics within the framework
of the mean-field approximation,
which describes the system as a gas of independent quasi-particles
endowed with the time-dependent effective mass obtained self-consistently
as indicated above (\ref{mu}).
In particular,
one may consider the evolution of a single pion mode $\k$
being subjected to the calculated effective mass $\mu_\pi(t)$.
This problem was studied recently in Ref.~\cite{JR:HIP}
and it was found that the result can be conveniently expressed
by means of the number enhancement coefficient $X_N$.
This quantity expresses by how much the occupancy of the given mode
is enhanced as a consequence of the prescribed time dependence of its mass.
Figure \ref{f:X} shows the resulting enhancement coefficient
as a function of the pion momentum,
obtained for the four cases depicted in Fig.~\ref{f:mass}.

On the basis of this schematic analysis,
one would expect a significant enhancement for the soft pion modes.
It appears that the effect is largest for the intermediate values of $T_0$,
as might have already been expected from the behavior seen
in Fig.~\ref{f:phaseBox}.
Moreover,
a careful inspection of the results reveals that $X_N(p)$
has two components:
a steady decrease resulting from the overall decay of $\mu_\pi(t)$
and a bump caused by the parametric amplification of modes
matching the frequency of the oscillatory part of $\mu_\pi(t)$,
with the former being by far the dominant component.
(The location at the bump at $200-250~\MeV/c$ corresponds to a pion
energy of about half the $\sigma$ mass,
as should be expected for the parametric resonance effect
\cite{Boyanovsky:PRD56b,KKRW,JR:HIP}.)

\subsection{Analysis of the final state}

As noted above,
the continual longitudinal expansion causes
the local field fluctuations to subside steadily in the course of time.
For the idealized one-dimensional expansion characteristic of Bjorken matter,
$\dphi^2$ decreases as $1/\tau$ for large times
(and faster than that when a system with a finite transverse cross section
is considered, as in the next section,
due to the additional dilution caused by the transverse expansion).
Therefore, after a sufficiently long time
the system will have decoupled into independently evolving modes
of ever decreasing field amplitude.
In this asymptotic regime,
it is possible to extract well defined values of the observables
by suitable analysis of the field configuration.
It is useful to note that for large times
the longitudinal velocity of a given part of the system
is given by $v_z\to z/t=\tanh(\eta)$.
Thus, in that limit,
the coordinate $\eta$ equals the rapidity ${\sf y}=\tanh^{-1}(v_z)$.

When analyzing the asymptotic field in the Bjorken expansion scenario,
it is natural to perform a Fourier transformation in the transverse plane,
\begin{eqnarray}\label{phik}
\pphi_\k(\eta) &=& 
\int{d^2\x\over \Omega_\perp}\ \pphi(\x,\eta)\ \rme^{-i\k\cdot\xscript}\ ,\\
\label{psik}
\ppsi_\k(\eta) &=& 
\int{d^2\x\over \Omega_\perp}\ \ppsi(\x,\eta)\ \rme^{-i\k\cdot\xscript}\ ,
\end{eqnarray}
where $\k=({k}_x,{k}_y)$ denotes the transverse wave vector and
$\Omega_\perp=L_xL_y$ denotes the transverse area of the spatial lattice.
In the present study,
we are only interested in observables based on
the three pion components of the chiral field.
Moreover,
while the dynamical treatment of these components
is carried out in a cartesian representation,
$\bold{\pi}=(\pi_1,\pi_2,\pi_3)$,
the observations are made in the spherical representation,
$\bold{\pi}=(\pi_-,\pi_0,\pi_+)$,
where the individual components represent the charge states of the pion,
\begin{equation}
\pi_\pm\ =\ {1\over\sqrt{2}}[\pi_1\pm i\pi_2]\ ,\hspace{2em} \pi_0\ =\ \pi_3\ .
\end{equation} 

The present treatment yields the time evolution of the classical
pion field, $\bold{\pi}(x,y,\eta,\tau)$, as described above.
In order to make contact with the physical observation,
it is necessary to transcribe the emerging field
into a quantal many-body state.
For this purpose,
we associate the asymptotic free field with a coherent state given by
\begin{eqnarray}\label{statek}
|\bold{\chi}\rangle &=& \exp\left\{ \sum_\k \int d\eta 
\left[\bold{\chi}_\k(\eta)\cdot\bold{a}_\k^\dagger(\eta)
-\bold{\chi}_\k^*(\eta)\cdot\bold{a}_\k(\eta)\right]\right\} |0\rangle\\
\label{statex}
&=& \exp\left\{ \int d^2\x \int d\eta 
\left[\bold{\chi}(\x,\eta)\cdot\bold{a}^\dagger(\x,\eta)
-\bold{\chi}^*(\x,\eta)\cdot\bold{a}(\x,\eta)\right]\right\} |0\rangle\ .
\end{eqnarray}
Here $a^{(j)}(\x,\eta)$ annihilates a pion of type $j$ 
located at $(\x,\eta)$.
It satisfies the usual commutation relation,
\begin{equation}
[a^{(j')}(\x',\eta'),a^{(j)}(\x,\eta)^\dagger]\ =\
\delta_{j'j}\ \delta^{(2)}(\x'-\x)\ \delta(\eta'-\eta)\ .
\end{equation}
Correspondingly,
$a_\k^{(j)}(\eta)$ annihilates a pion of type $j$ 
located longitudinally at $\eta$ and having the transverse wave vector $\k$,
with $[a_{\k'}^{(j')}(\eta'),a_\k^{(j)}(\eta)^\dagger]=
\delta_{j'j}\delta_{\k'\k}\delta(\eta'-\eta)$.

The coefficients in the expression (\ref{statek}) for the coherent state
are given simply in terms of the Fourier expansion coefficients in
Eqs.~(\ref{phik}-\ref{psik}),
\begin{equation}\label{chik}
\cchi_\k(\eta)\ =\ \sqrt{\Omega_\perp}
\left[ \sqrt{m_k\over2}\pphi_\k(\eta)
        +{i\over\sqrt{2m_k}}\ppsi_\k(\eta)\right]\ ,
\end{equation} 
where the transverse mass $m_k$ is given by
$m_k^2=m_\pi^2+k_x^2+k_y^2$.
The corresponding Fourier expansion yields the complex isovector field
appearing in the representation (\ref{statex}) of the coherent state,
\begin{equation}\label{chix}
\bold{\chi}(\x,\eta)\ \equiv\ {1\over\sqrt{\Omega_\perp}}
\sum_\k \bold{\chi}_\k(\eta)\ \rme^{i\k\cdot\xscript}\ .
\end{equation}
It presents a convenient encoding of the pion field configuration,
\ie\ the local field strength $\bold{\pi}(\x,\eta)$ and its time derivative.
We note that the coefficient $\cchi_\k(\eta)$ is represented by
the basic annihilation operator $\bold{a}_\k(\eta)$,
while $\cchi(\x,\eta)$ is represented by $\bold{a}(\x,\eta)$.

The coherent state $|\cchi\rangle$ has a number of convenient properties.
Most importantly,
it is an eigenstate of the annihilation operator,
\begin{eqnarray}
\bold{a}_\k(\eta)|\bold{\chi}\rangle &=& \cchi_\k(\eta)|\cchi\rangle\ ,\\
\bold{a}(\x,\eta)|\bold{\chi}\rangle &=& \cchi(\x,\eta)|\cchi\rangle\ ,
\end{eqnarray}
with the eigenvalues being the corresponding $\cchi$ coefficients.
The two-point density matrix for the state $|\bold{\chi}\rangle$
is then given simply in terms of the complex $\bold{\chi}$ field,
\begin{equation}
\langle\bold{\chi}| \bold{a}^\dagger(\x,\eta) 
\bold{a}(\x',\eta') |\bold{\chi}\rangle\
=\ \bold{\chi}^*(\x,\eta)\ \bold{\chi}(\x',\eta')\ .
\end{equation}
Thus the local density of pions with isospin component $j$ is given by
the corresponding diagonal element,
$d^3N^{(j)}/d^2\x d\eta=|\chi^{(j)}(\x,\eta)|^2$.
Furthermore,
the mean number of such pions emerging with transverse wave vector $\k$ 
and a rapidity $\sf y$ in the interval $({\sf y}_1,{\sf y}_2)$ is then given by
\begin{equation}\label{Nave}
{\bar n}_\k^{(j)}({\sf y}_1,{\sf y}_2)\ =\ 
\int_{{\sf y}_1}^{{\sf y}_2}d\eta\ |\chi_\k^{(j)}(\eta)|^2\ ,
\end{equation}
where we have utilized the asymptotic equivalence of 
the longitudinal coordinate $\eta$ and the rapidity $\sf y$.

In the numerical treatment,
the cartesian lattice has spacings of 
$0.2~\fm$ in the transverse plane and $\Delta\eta=0.2$ longitudinally.
Thus the numerical resolution in the longitudinal direction
corresponds to a rapidity {\em slice} of width $\Delta {\sf y}=0.2$.
In the analysis of the results,
we consider rapidity intervals $({\sf y}_1,{\sf y}_2)$ that have unit length,
corresponding to a physical rapidity resolution of one.
The corresponding sources are thus made up of five contiguous rapidity slices
and we shall refer to them as {\em lumps}.

\subsection{Transverse spectra}

The transverse spectrum of the emerging pions is conveniently
expressed as the invariant differential yield,
$Ed^3N/d^3\p=d^3N/d^2\k d{\sf y}$.
Due to the boost invariance of the scenario,
the resulting inclusive yield does not depend on the rapidity $\sf y$.
Moreover,
because of the rotational invariance in isospace,
it is the same for all three pion states.
Finally,
the overall axial symmetry of the geometry
guarantees that the resulting inclusive observables have azimuthal invariance.
Therefore,
the transverse spectra depend only on the transverse kinetic energy,
$E_k=m_k-m_\pi$.

In order to have a useful reference for judging the calculated spectra,
we introduce the equivalent temperature $T_{\rm BE}$
which is determined by matching the calculated spectrum
with a corresponding Bose-Einstein equilibrium form.
The energy per nucleon determines the temperature,
\begin{equation}
\epsilon\ \equiv\ {E\over N}\ =\ 
        {\sum_\k m_k f_k        \over   \sum_\k f_k }\ ,\hspace{2em}
f_k\ =\ [\rme^{m_k/T_{\rm BE}}-1]^{-1}\ ,
\end{equation}
and the appropriate normalization
is determined  subsequently by the actual number of particles $N$.
This fit is made by considering only a certain energy interval,
namely $200<E_k<1000$~\MeV.
The lower cutoff is made in order to permit the extraction
of the enhancement of soft pions that is expected to occur.
The exclusion of the high momenta is made for numerical convenience:
the high-momentum modes  (which have only a very small average occupancy)
are not treated with great accuracy in the present study
as they are computer-demanding but have a negligible effect on the results
of present interest.
(It has of course been checked that an upwards extension
of the upper cutoff on $E_k$,
with a corresponding better treatment of the high momentum modes,
has no import on the results extracted.)

\begin{figure}
~\vspace{180mm}
\includegraphics{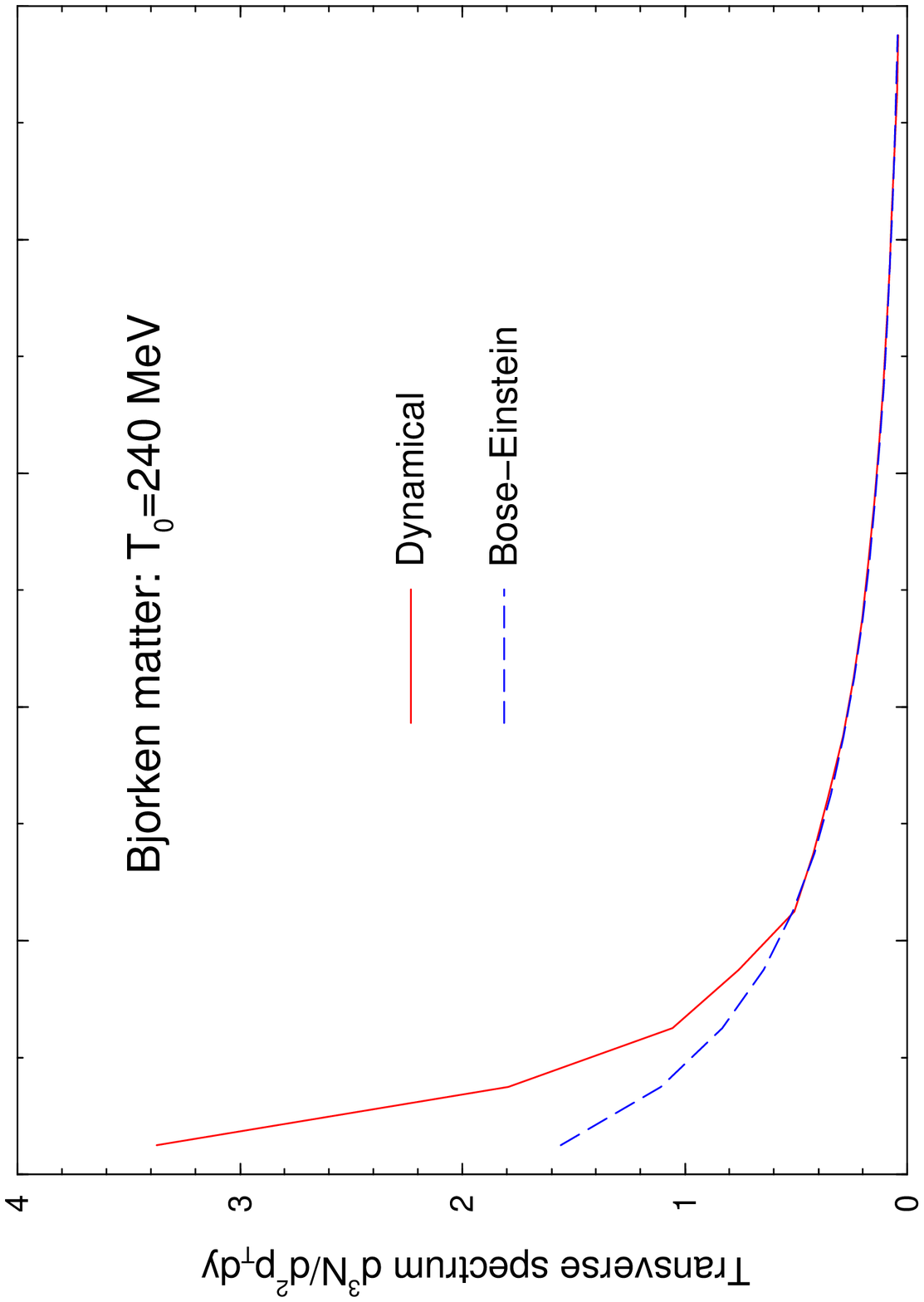}
\includegraphics{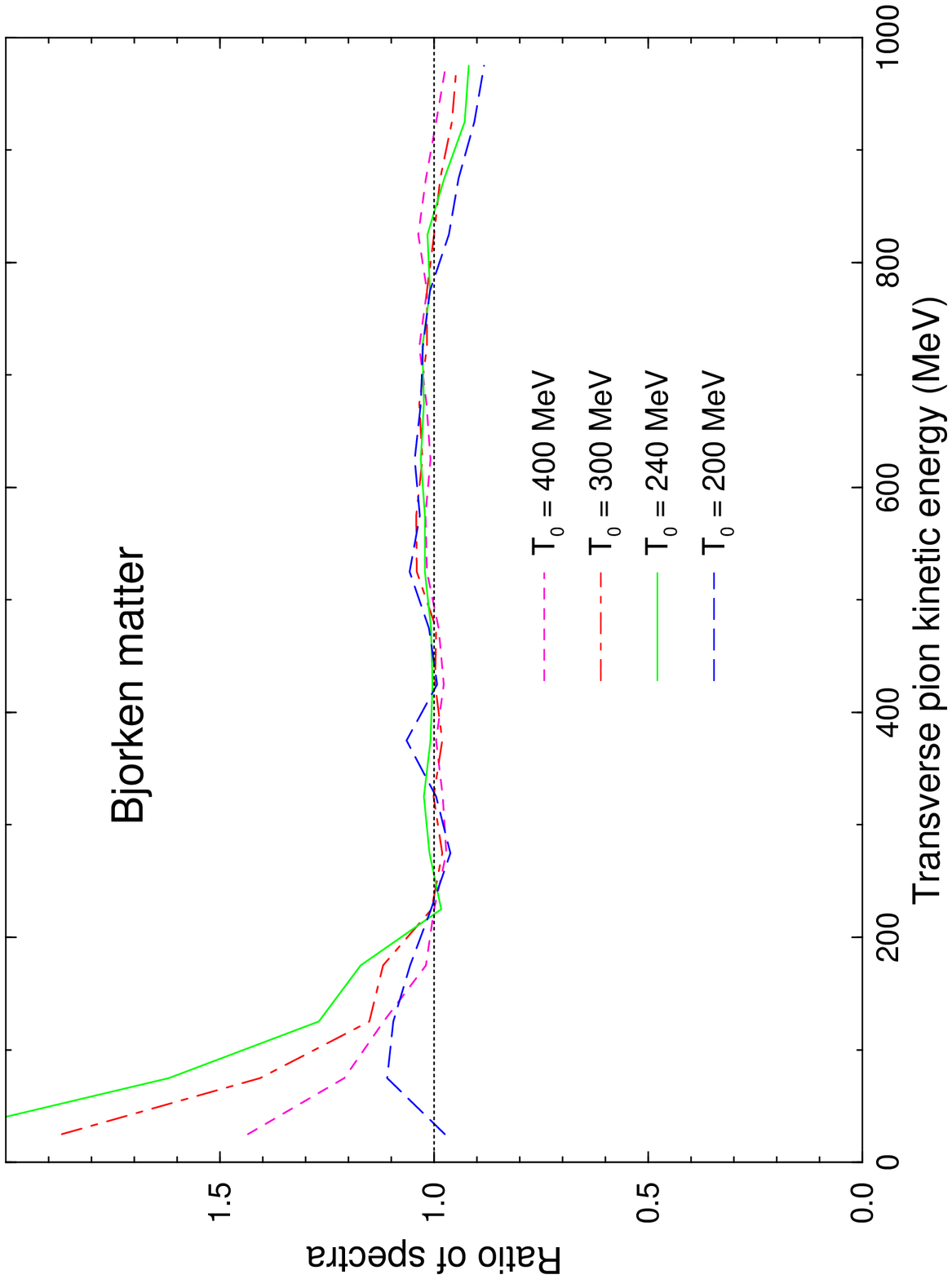}
\caption{Transverse spectrum for Bjorken matter.}
\label{f:spectrumBox}
{\small
The top panel shows the final transverse pion spectrum,
$d^3N/d^2\p_\perp d{\sf y}$,
and the associated equilibrium spectrum
obtained by fitting the result for the longitudinally expanding Bjorken box
with a Bose-Einstein form within the energy interval 200-1000 MeV,
for the initial temperature $T_0=240~\MeV$.
The bottom panel shows the ratio between
the dynamically generated spectrum and the fitted equilibrium form,
for four different initial temperatures $T_0$.
}
\end{figure}

As an example of the outcome of this procedure,
the top panel of Fig.~\ref{f:spectrumBox} shows the transverse spectrum
extracted from an ensemble of simulation events having $T_0=240~\MeV$,
together with the corresponding Bose-Einstein fit.
The fit is seen to be quite good for the region $E_k>200~\MeV$
within which it has been determined.
However,
below that region the dynamical result is significantly larger 
than the corresponding extrapolation of the fit.
In order to better bring out this enhancement of the soft pions,
we prefer to divide the calculated spectral profile
by the corresponding Bose-Einstein fit,
thus getting curves that are approximately unity above 200~MeV.
The bottom panel of Fig.~\ref{f:spectrumBox}
shows the resulting relative transverse spectra
for a number of initial temperatures $T_0$.
It is evident that the spectra are to a good approximation
of equilibrium form above 200~\MeV.
(The slight shortfall at the higher end of the interval
is due to the numerical cutoff in momentum space
and it has no import on our present discussion,
as discussed above.)
Furthermore, equally evident,
there is a significant enhancement of the soft pions
over a range of $T_0$ values,
with the effect being largest in the transition temperature region
where it reaches up to a factor of two for the softest pions.
(It should be noted that the relatively sudden onset of the enhancement
below 200~MeV precludes the possibility of obtaining fits 
of comparable quality by employing an adjustable chemical potential.)

Such a spectral enhancement would probably be within experimental reach.
However,
since the above results were obtained for idealized scenarios
of Bjorken matter,
they may well overestimate the effect.
It is therefore important to ascertain to what degree this enhancement
will persist as an observable signal
after the complications of a finite geometry have been included.

\section{Bjorken rod}

\begin{figure}
~\vspace{100mm}
\includegraphics{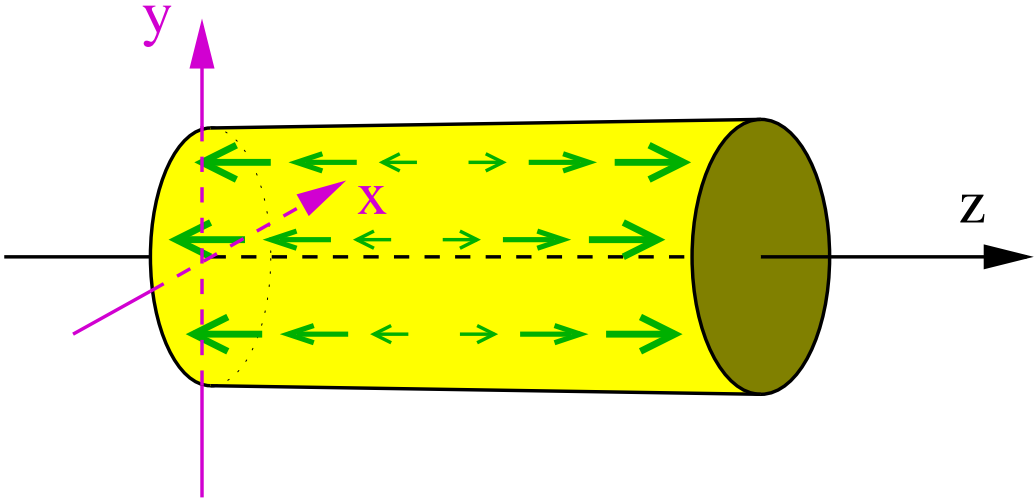}
\caption{The initial geometry of the Bjorken rod.}
\label{f:cyl}
{\small
The initial field configuration describes a rod-like system
subject to a longitudinal scaling expansion of the Bjorken type,
for which the local boost rapidity is given by ${\sf y}=z/\tau$.
}
\end{figure}

We now turn to the study of a more refined scenario
in which the system has a finite extension in the transverse plane.
Specifically,
preserving the longitudinal scaling expansion,
we shall replace the matter scenario by a rod-like geometry
for which the local environment changes
from that of hot Bjorken matter in the bulk
to that of vacuum outside
through a smooth surface region with a circular cross section.
For convenience, we shall refer to such a system as a {\em Bjorken rod}.
It is illustrated in Fig.~\ref{f:cyl}.

The Bjorken rod combines two features exhibited
 by the systems produced in real nuclear collisions:
a rapid longitudinal expansion
and a finite transverse extension.
This latter feature implies the presence of a surface region 
through which the order parameter changes
from its initially small value in the hot interior
to its vacuum value outside.
Because it is both relatively realistic and reasonably tractable,
this model scenario has been widely employed in high-energy heavy-ion studies.
In the context of DCC studies,
it was employed already early on by Asakawa {\em et al.}\ \cite{AHW:PRL74}.
In that work,
the chiral field was taken to be constant in the longitudinal direction,
thus rendering the system effectively two-dimensional
(with a corresponding reduction of the pressure in the interior).
Furthermore,
the present study employs a more refined initialization procedure
which endows the system with more realistic features,
such as a finite correlation length and a surface region
through which the field changes smoothly from the hot interior to the
exterior vacuum.
Finally,
whereas the work reported in Ref.~\cite{AHW:PRL74} was mainly exploratory
and rather limited in scope,
the present study involves quantitative analyzes of the final state
in order to make contact with specific experimental observables.
It should be emphasized that the general conclusions drawn 
in that exploratory work are not affected
by the results of the present study.

\subsection{Preparation of the rod}

For the rod scenario,
we again employ a rectangular lattice in $(x,y,\eta)$
and impose periodic boundary conditions in all three directions.
In the transverse $xy$ plane,
the size of the box should be sufficient
to contain the expanding system at times up to the point
when the analysis of the final state is made;
when this is ensured,
the transverse boundary conditions are immaterial.
In the longitudinal $\eta$ direction,
the periodicity approximates conditions that would prevail
if the cylinder were infinitely long,
as in the idealized scenario originally considered by Bjorken \cite{Bjorken}.
Typically the box sides employed are $L_x=L_y=44\ \fm$,
which suffices for the systems considered,
and $L_\eta=8$,
corresponding to that many units of rapidity.

In preparing the system,
we wish the local conditions to correspond approximately to thermal equilibrium
at a temperature that decreases steadily from $T_0$ inside the bulk region
to zero in the vacuum outside.
The local temperature $T$ has a Saxon-Woods profile,
\begin{equation}\label{WS}
T(\rho)\ =\ T_0~[1+\rme^{(\rho-R_0)/w}]^{-1}\ ,
\end{equation}
where $\rho^2=x^2+y^2$.
For the surface width parameter we have used $w=0.8~\fm$ and
we show results obtained for ensembles of rods with $R_0$ equal to 6 or 10 fm.
We expect such values of $R_0$ to provide a lower bound on
the transverse extensions that may be reached
for central collisions of gold nuclei at RHIC energies
by the time the system has cooled to the specified value of $T_0$.
This feature is convenient,
since a system with a larger transverse extension
would be closer to the Bjorken matter scenario considered above 
(which is approached in the limit $R_0\to\infty$).

In order to prepare the initial field configuration for the rod,
we proceed at first in the same manner as for the preparation of the
matter scenario addressed above
and sample the field configuration $(\pphi_{\rm box},\ppsi_{\rm box})$ from
a thermal ensemble describing macroscopically uniform matter
within the overall box containing the calculational lattice.
This field configuration can be uniquely decomposed into its spatial average,
the order parameter, and the remainder, the fluctuating part of the field,
\begin{eqnarray}
\pphi_{\rm box}(\x,\eta)\ &=&\ \ul{\pphi}\ +\ \dpphi(\x,\eta)\ ,\\
\ppsi_{\rm box}(\x,\eta)\ &=&\ \ul{\ppsi}\ +\ \dppsi(\x,\eta)\ .
\end{eqnarray}
In order to obtain the field configuration
describing the initial state of the rod, 
$(\pphi_{\rm rod},\ppsi_{\rm rod})$,
we rescale the two parts based on the specified local temperature $T(\rho)$
and then recombine them into the desired initial conditions,
\begin{eqnarray}
\pphi_{\rm rod}(\x,\eta,\tau_0)\ &=&\ 
g(\rho)[\ul{\pphi}-\pphi_{\rm gs}]\, +\pphi_{\rm gs}\ 
+\ \bold{h}(\rho)\circ[\dpphi(\x,\eta)-\ul{\pphi}]\, +\ul{\pphi}\ ,\\
\ppsi_{\rm rod}(\x,\eta,\tau_0)\ &=&\ 
g(\rho)[\ul{\ppsi}-\ppsi_{\rm gs}]+\ppsi_{\rm gs}\
+\ \bold{h}(\rho)\circ[\dppsi(\x,\eta)-\ul{\ppsi}]+\ul{\ppsi}\ ,
\end{eqnarray}
where $\pphi_{\rm gs}=(f_\pi,\bold{0})$
and $\ppsi_{\rm gs}=(0,\bold{0})$ are the vacuum values.
The scaling coefficient for the order parameter
is obtained from the magnitudes of the corresponding order parameters,
$g(\rho)= [\phi_0(T(\rho))-f_\pi]/[\phi_0(T_0)-f_\pi]$,
where $\phi_0(T)$ denotes the magnitude of the order parameter
in thermal equilibrium at the specified temperature $T$.
The scaling coefficient for the fluctuations
depends on the particular chiral component,
so $\bold{h}$ is a diagonal O(4) tensor
with the following elements,
\begin{equation}
h_\sigma(\rho)\ =\ 
\left({\delta\sigma^2_T \over \delta\sigma^2_{T_0}}\right)^{1\over2}\ ,\,\,\
h_\pi(\rho)\ =\ 
\left({\delta\pi^2_T \over \delta\pi^2_{T_0}}\right)^{1\over2}\ .
\end{equation}
and $\circ$ denotes the corresponding O(4) scalar product.
Here $\delta\sigma^2_T$ denotes the thermal equilibrium value
of the fluctuation in the $\sigma$ component of the chiral field
and $\delta\pi^2_T$ is the thermal fluctuation
in (any) one of the pion components.
These are calculated by the semi-classical mean-field method \cite{JR:PRD}.
(It should perhaps be noted that in principle one needs to take account
of the fact that generally the order parameter $\ul{\pphi}$
is not fully aligned with the O(4) $\sigma$ direction
and hence the $\sigma$ and $\pi$ field fluctuations are not independent.
Rather, it is the fluctuations parallel and perpendicular to
the order parameter, $\delta\phi_\parallel$ and $\dpphi_\perp$, that decouple
(\ie\ their quasiparticle mass tensor is diagonal).
However,
for the quite large systems considered here,
$\ul{\pphi}$ is practically fully aligned in the $\sigma$ direction
and so we do not need to be concerned with that distinction
in our discussion.
Nevertheless, for the sake of generality,
the numerical treatment performs the proper O(4) diagonalization
and rotation when the field preparation is done \cite{JR:PRD}.)

By proceeding in the manner described above,
we ensure that the local environment,
as characterized by the order parameter and the field fluctuations,
reflects approximately thermal equilibrium in matter held at the
local temperature $T$, which decreases from its bulk value $T_0$ to zero
according to the prescribed radial profile (\ref{WS}).

\begin{figure}
~\vspace{180mm}
\includegraphics{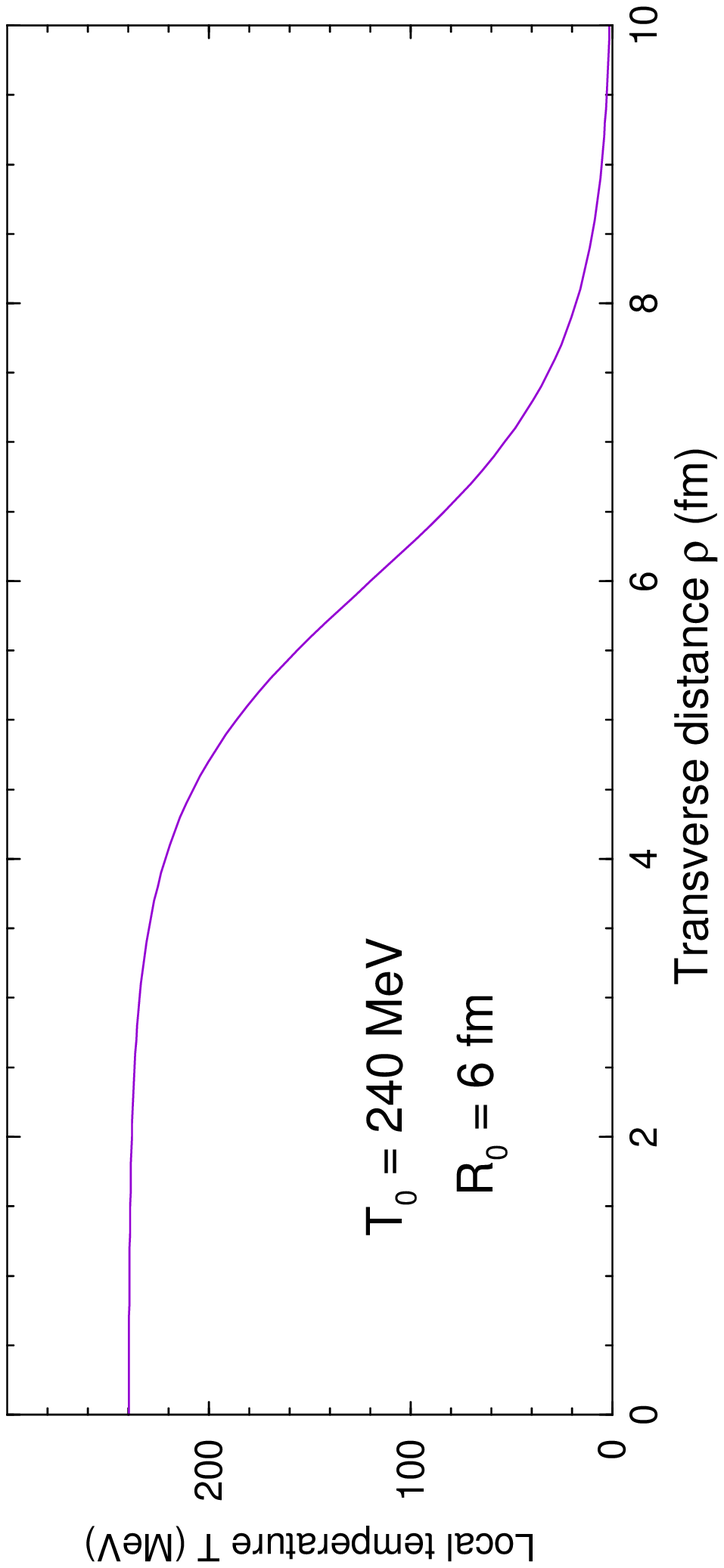}
\includegraphics{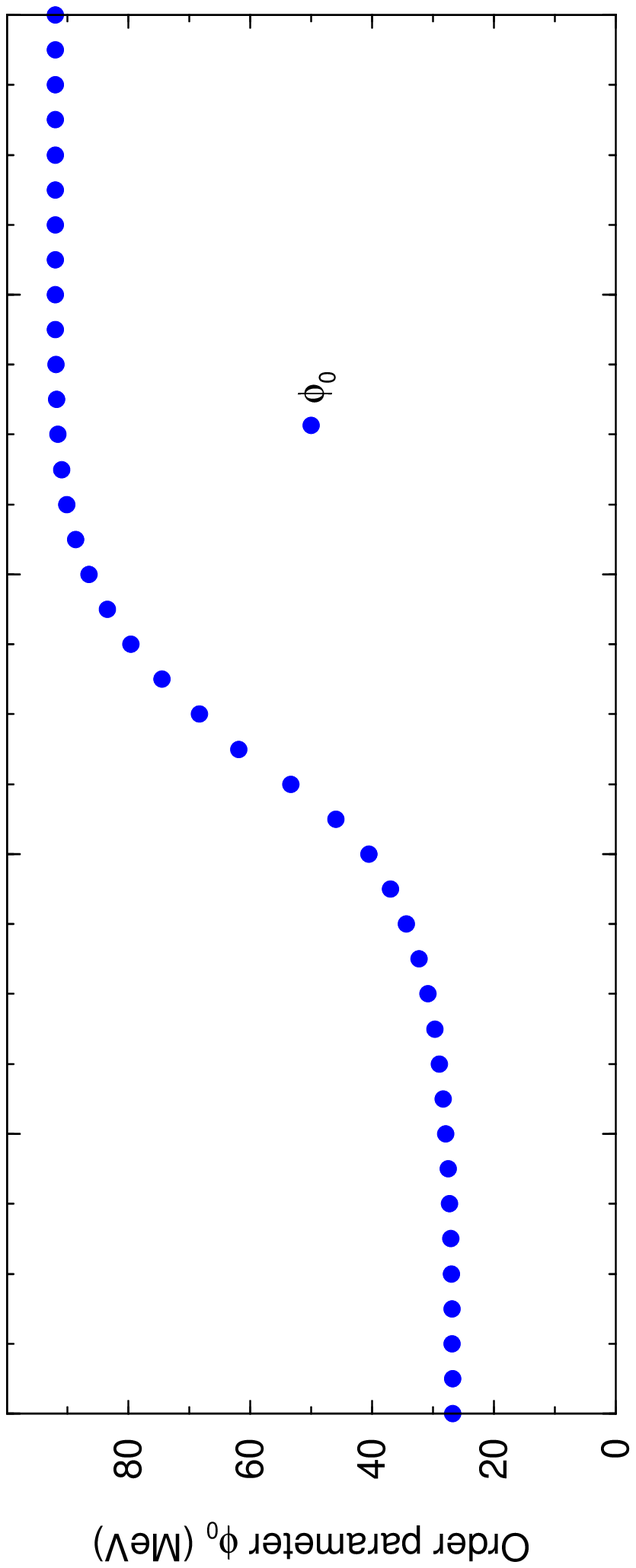}
\includegraphics{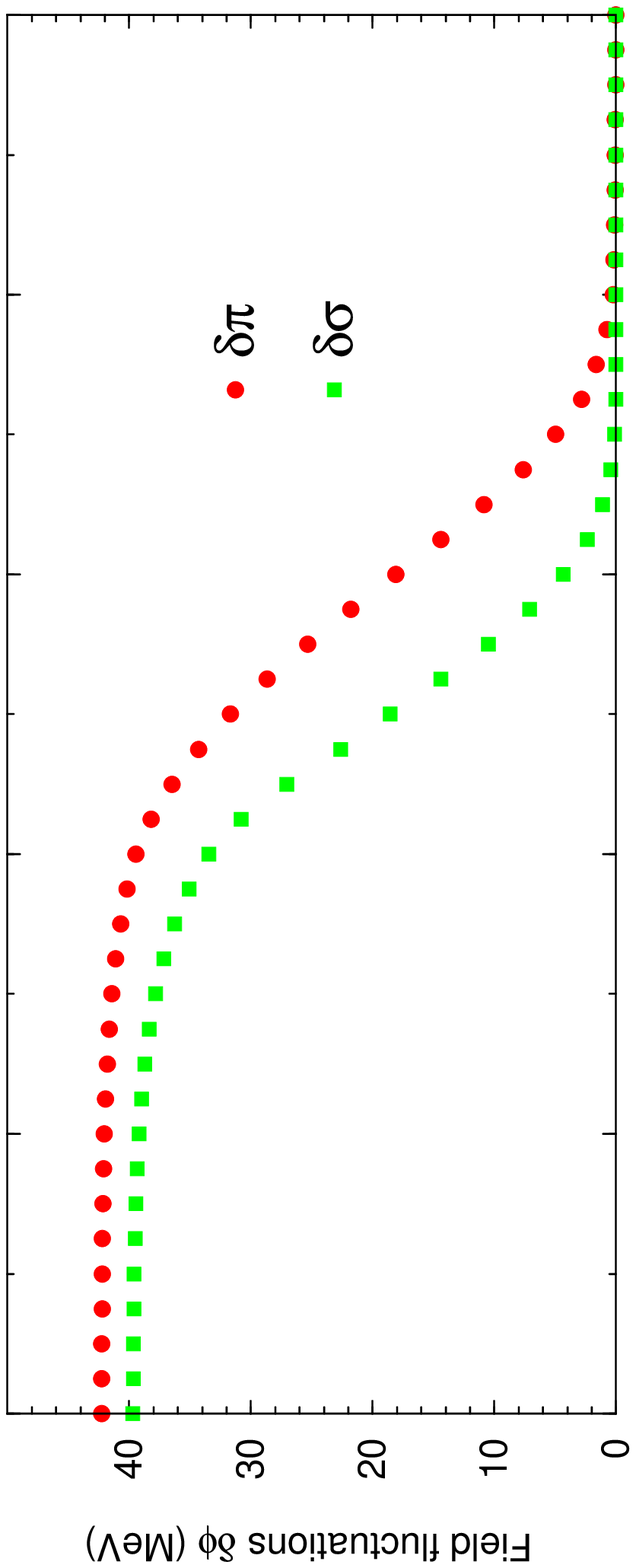}
\includegraphics{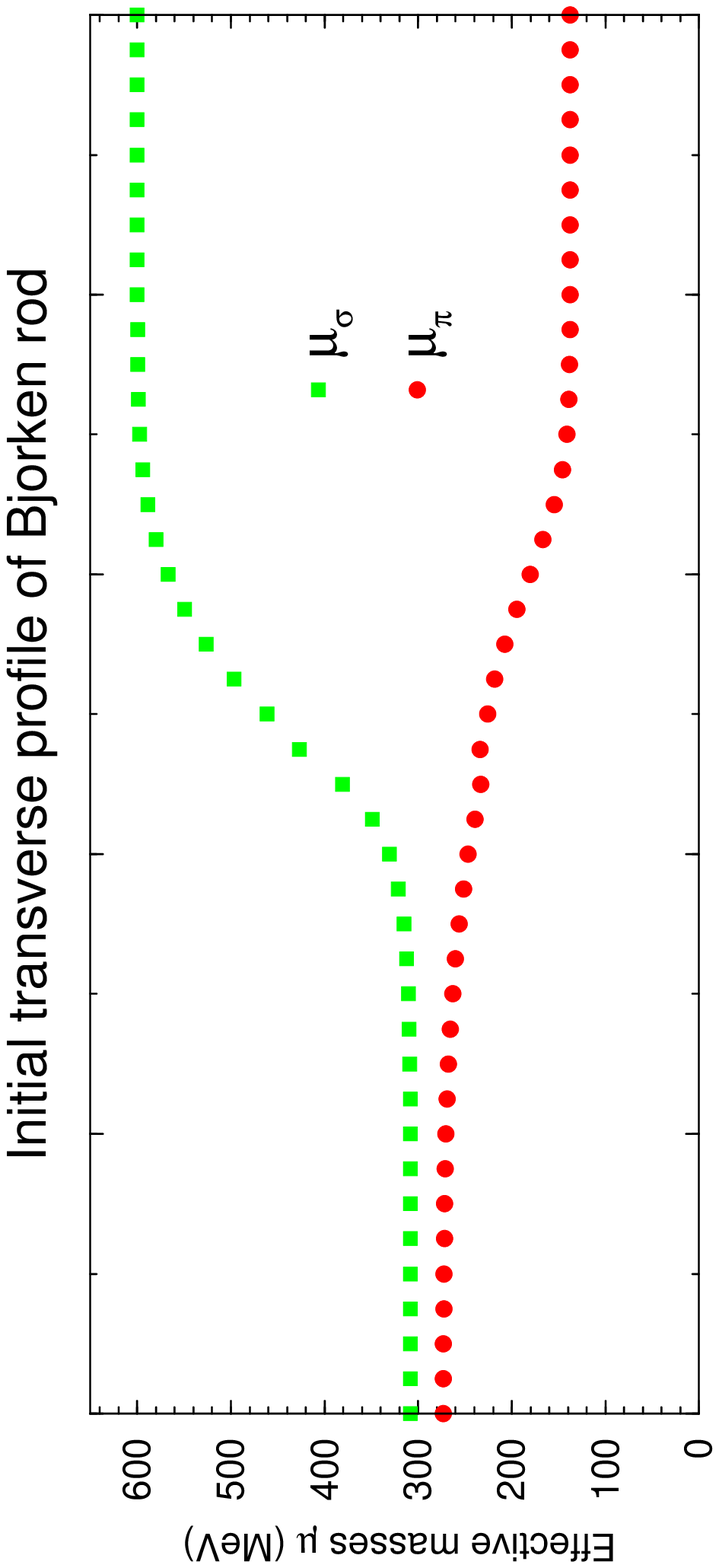}
\caption{Initial rod profiles.}
\label{f:profiles}
{\small
The transverse profiles of
the local temperature $T$,
the order parameter $\phi_0$,
the mean fluctuation of the field in a given O(4) direction,
$\delta\sigma$ and $\delta\pi$,
and the corresponding effective masses $\mu_\sigma$ and $\mu_\pi$,
for a rod with a radius of $R_0=6~\fm$ 
and a central temperature of $T_0=240~\MeV$.
}\end{figure}

The initial system is illustrated in Fig.~\ref{f:profiles}
for $T_0=240~\MeV$.
It shows the radial dependence of a number of key quantities:
the local temperature $T$,
the order parameter $\phi_0$,
the field fluctuations $\delta\sigma$ and $\delta\pi$,
and the effective masses $\mu_\sigma$ and $\mu_\pi$.
Since the local temperature drops steadily 
as a function of the transverse distance $\rho$,
moving out along the abscissa corresponds to reducing the temperature
(though not at a steady rate).
As one moves out through the surface,
the order parameter increases steadily 
from its reduced value ($\approx27~\MeV$) in the hot bulk region
towards its vacuum value $f_\pi$ ($=92~\MeV$),
and the local thermal fluctuations drop correspondingly towards zero.
(Since generally $\mu_\sigma>\mu_\pi$,
the fluctuations along a pion direction 
exceed those in the $\sigma$ direction.)
The resulting profiles of the effective masses
also reflect their temperature dependences:
Starting from nearly degenerate values ($\approx300~\MeV$) in the hot interior,
where chiral symmetry is approximately restored,
$\mu_\pi$ and $\mu_\sigma$ diverge steadily towards their free values of
$138~\MeV$ and $600~\MeV$, respectively.
For higher values of the central temperature $T_0$,
the central value of the order parameter is smaller
and the effective masses are larger (and even closer in value)
and $\mu_\sigma$ will in fact exhibit a dip in the surface region
as the local temperature passes through the critical region \cite{JR:PRD,Tchi}.

\subsection{Rod dynamics}

\begin{figure}
~\vspace{180mm}
\includegraphics{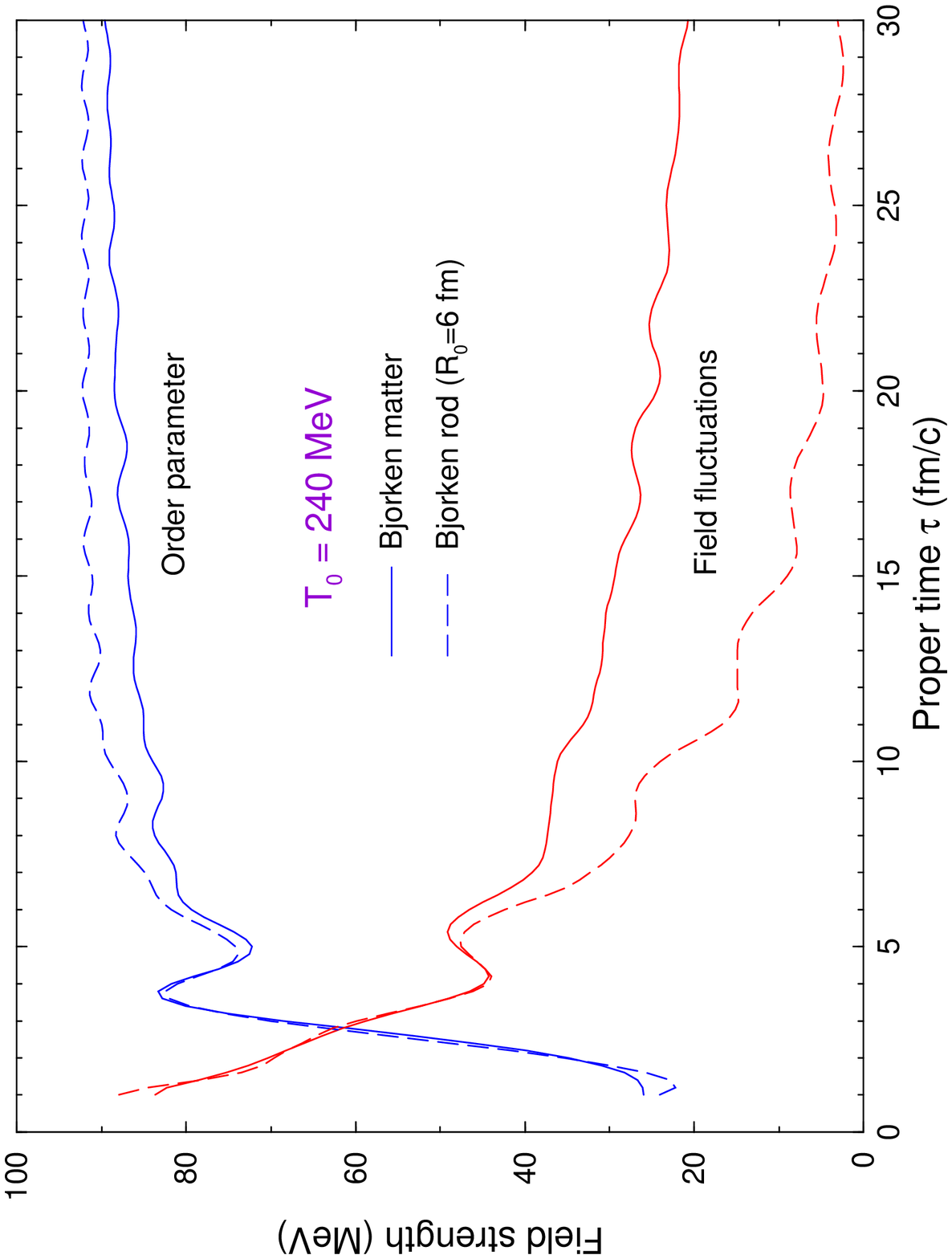}
\caption{Time evolution of the field.}
\label{f:time240}
{\small
The time dependence of $\phi_0$,
the magnitude of the order parameter (increasing curves),
and $\delta\phi$,
the associated dispersion of the field fluctuations (decreasing curves),
in both Bjorken matter prepared with $T_0=240~\MeV$ (solid curves)
and the corresponding Bjorken rod with $R_0=6~\fm$ (dashed curves).
The information for the rod has been obtained by averaging
over a hollow cylindrical volume, $1~\fm< \rho < 3~\fm$.
(The region very close to the symmetry axis has been excluded
in order to avoid the enhanced fluctuations in the center
caused by the strict axial symmetry of the geometry.)
}
\end{figure}

After the preparation of the field configuration of the rod,
the dynamical propagation is readily obtained by solving the
field equation (\ref{EoM}) as was done for the matter configurations.
Since the system is no longer macroscopically uniform,
it is more complicated to discuss,
but it is especially instructive to see how the bulk region develops.
For this purpose,
we probe a circular region in the interior of the rod
and extract the corresponding values of the order parameter
and field fluctuations.
The resulting evolution of the order parameter and the field fluctuations
is illustrated in Fig.~\ref{f:time240}
for a rod with $T_0=240~\MeV$ and $R_0=6~\fm$.
The corresponding quantities for Bjorken matter prepared
with the same value of $T_0$ are also shown.
It is seen that the environment in the interior of the rod evolves
in a manner quantitatively very similar to that of the corresponding
matter scenario throughout the first complete oscillation of the
the order parameter, which takes about $4~\fm/c$.
At around that point in time,
the decompression wave has arrived from the rod surface
and the self-generated transverse expansion now extends over the entire
cross section of the rod.
As a consequence, the relaxation progresses faster
than it would in matter where this effect is absent.%
\footnote{
The quicker relaxation of the rod configuration is a computational advantage,
as it compensates somewhat for the larger lattice needed
to enable the rod to expand transversally.}

\begin{figure}
~\vspace{180mm}
\includegraphics{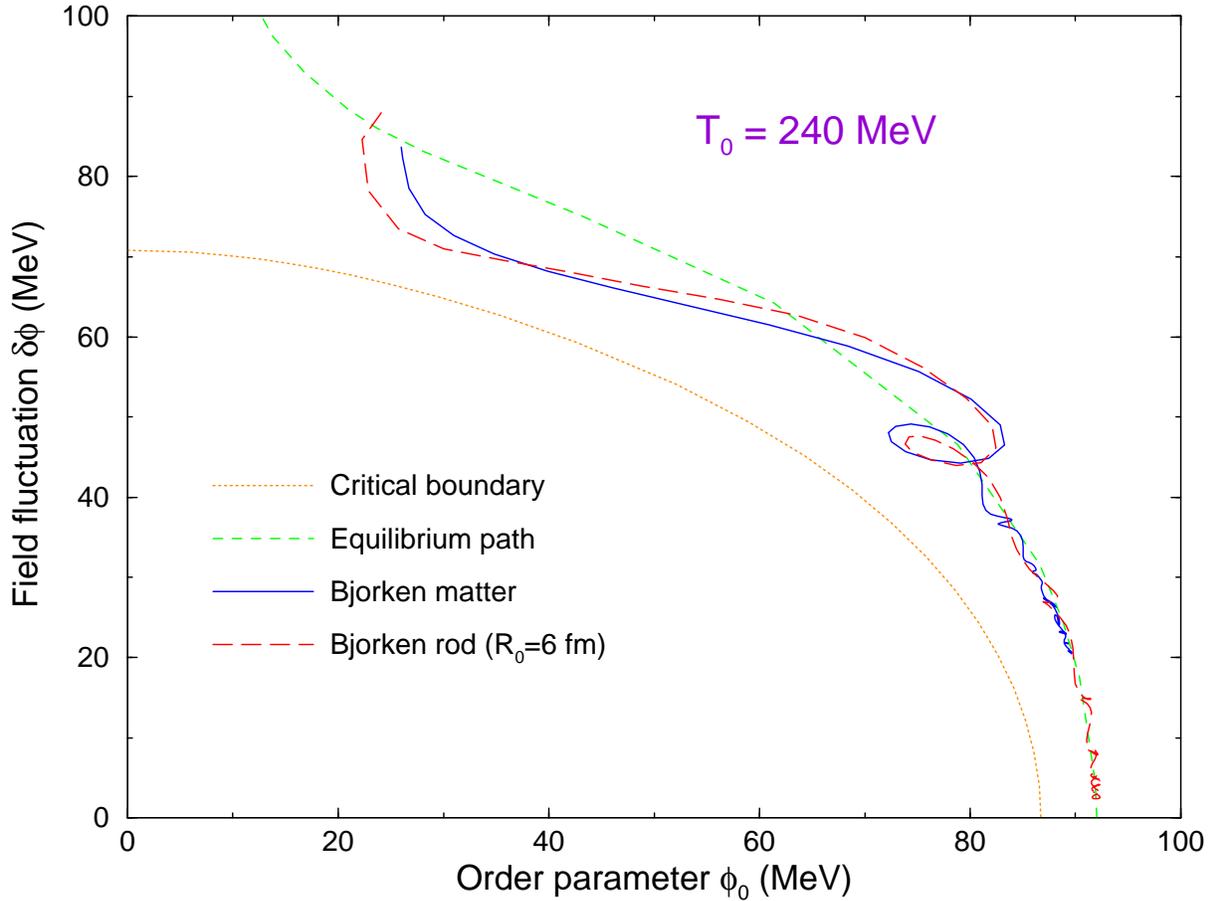}
\caption{Phase evolution of the rod.}
\label{f:phase240}
{\small
The phase evolution of the interior of the Bjorken rod
prepared with a bulk temperature of $T_0=240~\MeV$ 
and a radius of $R_0=6~\fm$ (dashed path, filled diamonds),
compared with the corresponding evolution of Bjorken matter
shown in Fig.~\ref{f:phaseBox} (solid path, open circles).
The equilibrium path is shown by the short-dashed curve
while the dotted curve delineates the region of instability
within which the field is supercritical.
The information for the rod has been obtained by averaging
over a hollow cylindrical volume, $1~\fm< \rho < 3~\fm$.
The symbols give the location of the paths 
at successive proper times $\tau=1,2,3,4,5,10,20,30~\fm/c$.
}
\end{figure}

This feature is also brought by the the corresponding dynamical paths
on the phase diagram, as shown in Fig.~\ref{f:phase240}.
The initial location of the point for the rod interior
differs slightly from that of the equilibrium matter
because of the finite size of the hollow cylindrical volume sampled.
But apart from this minor difference,
the rod path tracks the matter path closely for the first several time units,
as brought out by the proximity of the corresponding time
markers at $\tau=1,2,3,4,5~\fm/c$.
Beyond that time,
the paths begin to diverge,
as shown by the subsequent time markers at $\tau=10,20,30~\fm/c$.
From around $20~\fm/c$, or so, the rod interior is quite close
to the vacuum point,
whereas the environment in the matter scenario remains significantly excited
for a much longer time.

\begin{figure}
~\vspace{180mm}
\includegraphics{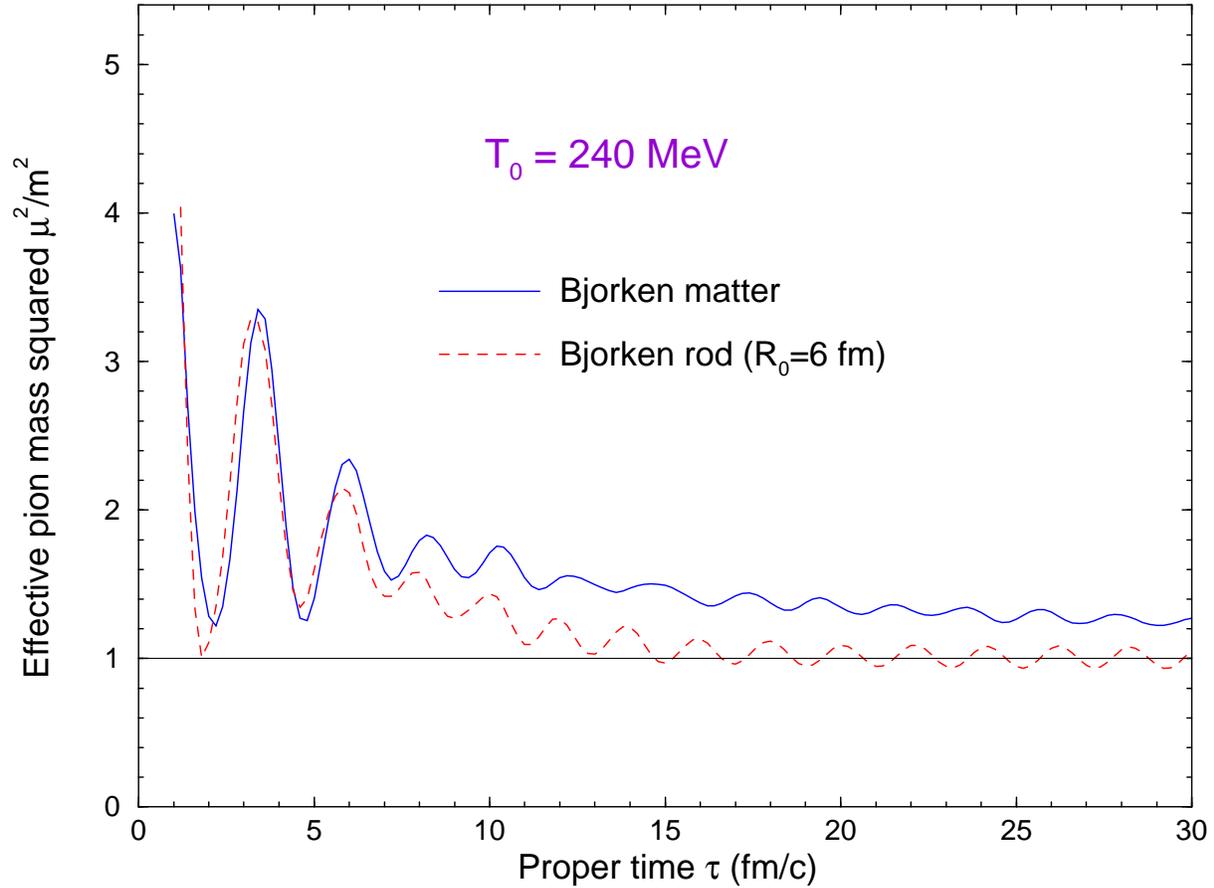}
\caption{Effective mass.}
\label{f:mass240}
{\small
The time evolution of the square of the effective pion mass, $\mu_\pi^2(t)$
(divided by the square of its free mass $m_\pi$)
for a Bjorken rod (dashed curve) prepared with $T_0=240~\MeV$ and $R_0=6~\fm$ 
as well as the corresponding result for Bjorken matter
taken from Fig.~\ref{f:mass} (solid curve).
The information for the rod has been obtained by averaging
over a hollow cylindrical volume, $1~\fm< \rho < 3~\fm$.
}
\end{figure}

The quicker relaxation of the field for the rod configuration
is also reflected in the behavior of the effective pion mass.
This is illustrated in Fig.~\ref{f:mass240} for the same case.
Again we see that through the first several time units
the effective mass in the interior of the rod
follows closely the evolution of the effective mass
extracted for the corresponding matter scenario,
but it then drops much faster towards the free value.
Furthermore, the oscillations in the effective mass
persist for quite a long time, thus making it possible to achieve
a significant degree of parametric amplification.

\subsection{Transverse spectra}

We now consider a number of specific observables
that are practically accessible in the analysis of actual experimental data.
First we consider the transverse spectra of the emerging pions,
$Ed^3N/d^3\p=d^3N/d^2\k d{\sf y}$.
As was the case for the Bjorken matter scenarios,
the transverse spectral shape is well approximated by an equilibrium form
above kinetic energies of $200~\MeV$ or so.
We therefore proceed in the same manner
and extract a corresponding equivalent temperature, $T_{\rm BE}$,
for each case considered.

The values obtained for $T_{\rm BE}$ are shown in Table~\ref{t:TBE}
for a range of initial bulk temperatures $T_0$
and a rod radius of $R_0=6~\fm$
(the values of $T_{\rm BE}$ are rather insensitive to $R_0$).
We note that they increase steadily with $T_0$,
as one would expect.
It might at first seem surprising that $T_{\rm BE}$,
the effective temperature characterizing the transverse spectrum
of the final pions,
can exceed $T_0$,
the specified bulk temperature of the initial state.
However,
it should be kept in mind that the quasipions in the bulk of the initial system
have an effective mass $\mu_\pi$ that exceeds the free value $m_\pi$
by hundreds of MeV,
due to the self-interaction of the thermally agitated field.
As the system expands and cools,
the interaction energy energy is converted primarily into kinetic energy
and hence the spectrum hardens.

\begin{figure}
~\vspace{180mm}
\includegraphics{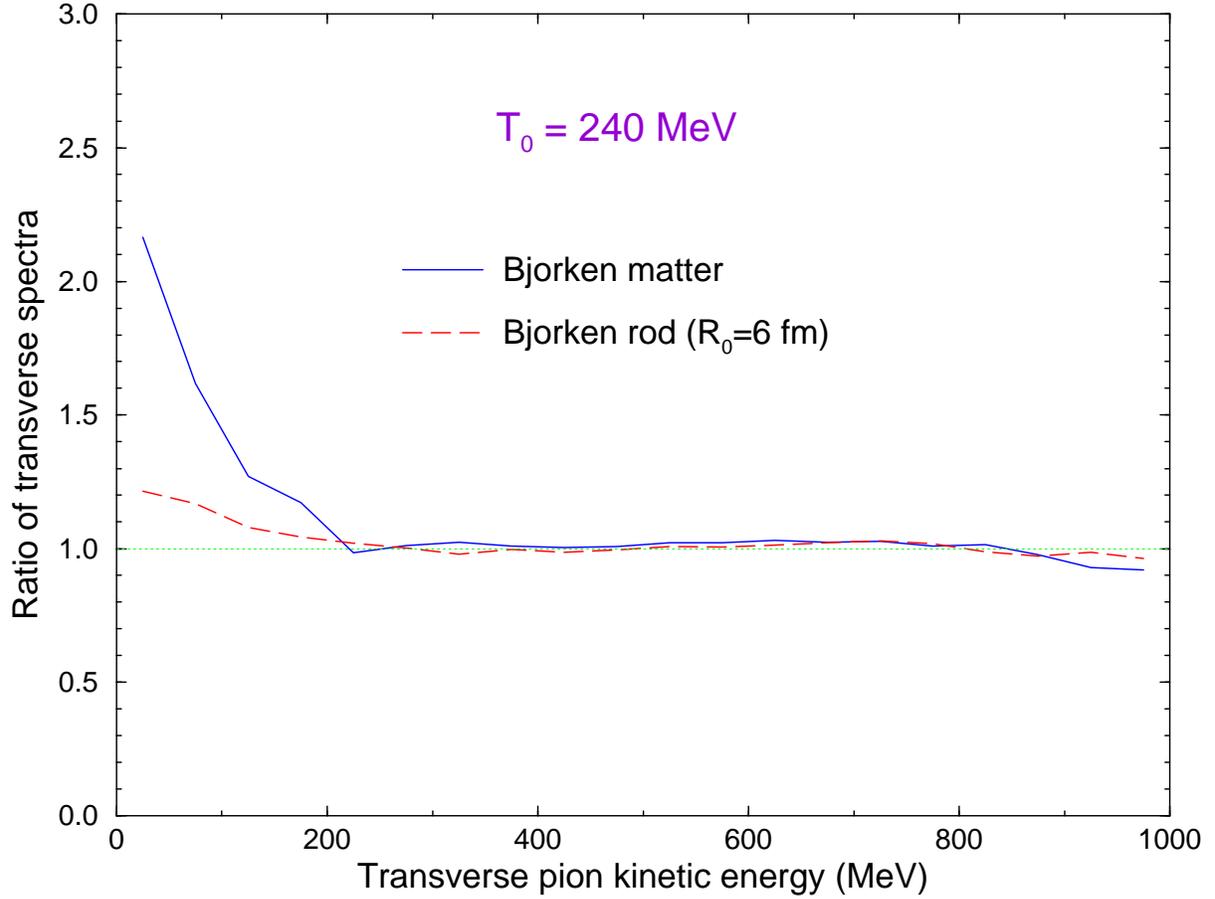}
\caption{Spectral shape for Bjorken rods.}
\label{f:ratioRod}
{\small
The ratio between the final transverse pion spectrum,
$d^3N/d^2\p_\perp dy$,
and the associated equilibrium spectrum
obtained by fitting the dynamical result
with a Bose-Einstein form within the energy interval 200-1000 MeV,
for a longitudinally expanding rod (dashed curve)
having an initial radius of $R_0=6~\fm$
and with an initial bulk temperature of $T_0=240~\MeV$,
as well as the corresponding result for Bjorken matter
(from Fig.~\ref{f:spectrumBox}).
}
\end{figure}

The resulting ratio between the dynamical spectrum
and the corresponding Bose-Einstein form
is displayed in Fig.~\ref{f:ratioRod} for $T_0=240~\MeV$,
together with the corresponding matter result.
While the quality of the fit for rods is as good as for matter,
the ensuing soft enhancements are generally smaller.
This is not surprising,
since the bulk conditions prevail in only part of the rod configuration and,
moreover, the transverse expansion tends to suppress the oscillatory
relaxation causing the enhancement.

\begin{figure}
~\vspace{180mm}
\includegraphics{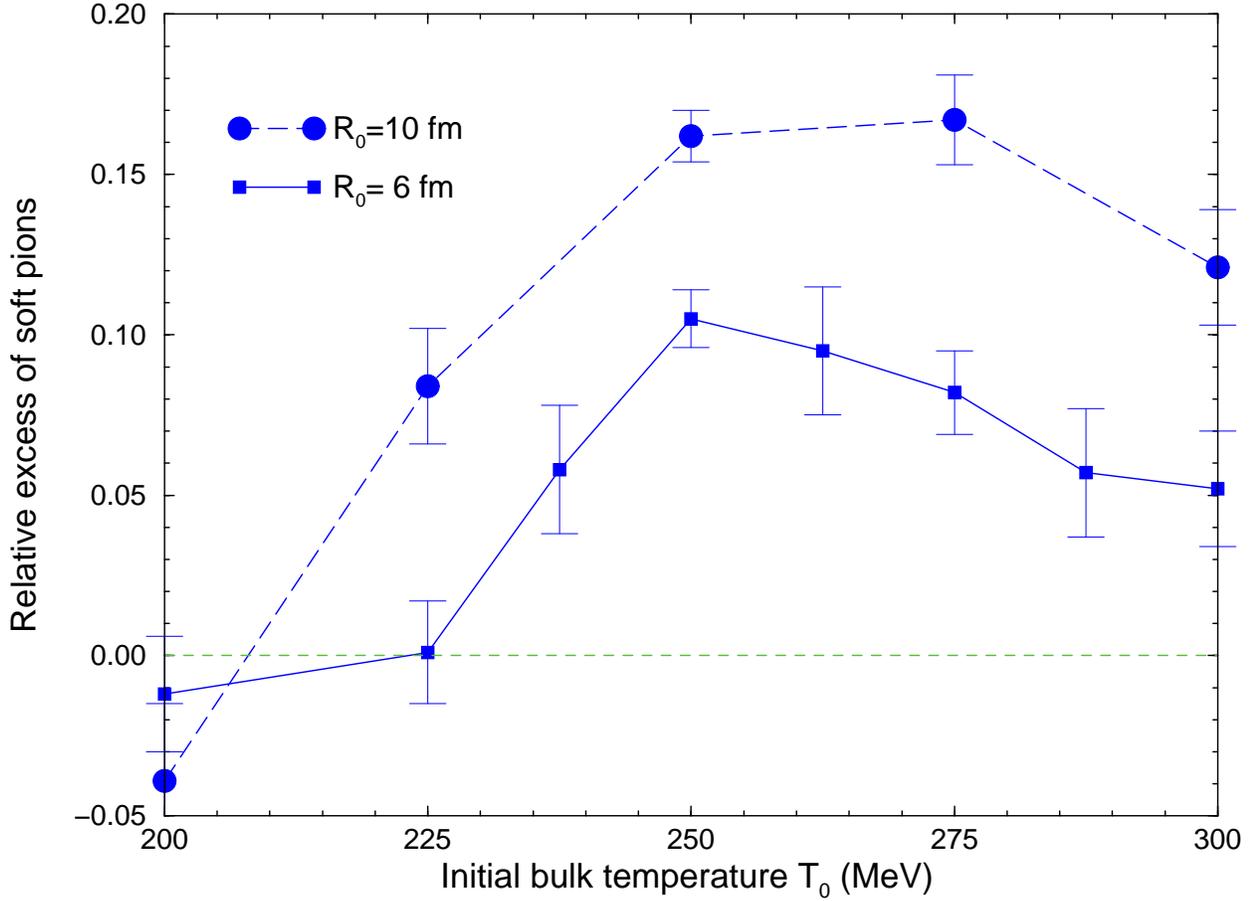}
\caption{Excess of soft pions for Bjorken rods.}
\label{f:excessRod}
{\small
The relative excess of pions for Bjorken rods
with initial radii $R_0$ equal to either 6 or 10 fm,
as a function of the initial bulk temperature $T_0$.
The excess has been obtained by subtracting the yield
of pions with kinetic energies below 200 MeV
from the corresponding equilibrium spectrum
obtained by fitting the dynamical result
with a Bose-Einstein form within 200-1000 MeV.
}
\end{figure}

Figure~\ref{f:excessRod} shows the relative excess of soft pions
as a function of the initial bulk temperature $T_0$
for Bjorken rods with $R_0$ equal to 6 and 10 fm.
We see that the effect begins to manifest itself
at temperatures above 200~MeV,
for which the bulk value of the order parameter is significantly reduced
from its vacuum value,
and it exhibits a broad maximum at roughly $T_0=240-280~\MeV$
before gradually subsiding at higher $T_0$.
The larger the rod radius $R_0$,
the larger the enhancement,
as one would expect,
since the system is then better approximated by bulk matter.
The excess of pions below 200~MeV is of the order of ten per cent.
Though significantly smaller than that obtained for infinite matter,
such enhancements are nevertheless still significant
and should be well within observable reach.

Since the excess pions reflected in the spectral enhancement
are produced by the coherent oscillation of the order parameter,
their properties are expected to differ from the regular pions.
Therefore,
in order to better bring this out,
it is useful to consider the soft and hard pions separately
in the following analyses.

\subsection{Multiplicity distributions}

We now turn to the discussion of the resulting multiplicity distributions.
By invoking the relation (\ref{Nave}),
it is straightforward to calculate the expected number of pions
in a given rapidity interval and within a specified transverse energy bin.
Of course,
this result varies from event to event,
since the field configurations vary at the microscopic level
as a result of the thermal fluctuations in the ensemble of initial states.

For a given value of the expected number $\bar{n}$,
the actual number of particles emerging is a stochastic variable $n$,
since the state described by a given field has no well-defined particle number.
Generally,
knowledge of the classical field alone does not allow a construction
of the underlying quantum many-particle state.
However,
in the present case where we have made the common assumption
that the final quantum state is of standard coherent form
(see Eq.~(\ref{statek})),
it is elementary to show that the associated multiplicity distribution
is of Poisson form.
Thus, the probability for obtaining a specified number of particles
with transverse momentum $\k$
within a given rapidity interval,
$n_\k^{(j)}({\sf y}_1,{\sf y}_2)=n$, is given by
\begin{equation}
P_{\bar n}(n)\ =\ {\bar{n}^n \over n!}\ \rme^{-\bar{n}}\ ,
\end{equation} 
where $\bar n=\bar{n}_\k^{(j)}({\sf y}_1,{\sf y}_2)$
is the given mean value (\ref{Nave}).
We can therefore readily pick the actual number of $j$-type pions
emerging from each particular rapidity slice with the transverse momentum $\k$.
The Poisson distributions form a closed algebra under convolution,
with the corresponding mean values being additive,
$P_a*P_b=P_{a+b}$.
Consequently,
the total number of particles emerging with a given transverse wave vector $\k$
from a given rapidity lump can be obtained equivalently by either
sampling the multiplicities for each of the constituent slices
from the corresponding individual Poisson distributions,
or by sampling the the total multiplicity $n_\k$ from the
corresponding combined Poisson distribution.
Different values of $\k$ can be combined in a similar manner,
for example to obtain the number of pions emerging within
a specific energy bin.
It is thus clear that
the resulting calculated statistical multiplicity distributions
do not depend on the specific computational procedure employed
for the stochastic sampling. 

\begin{figure}
~\vspace{160mm}
\includegraphics{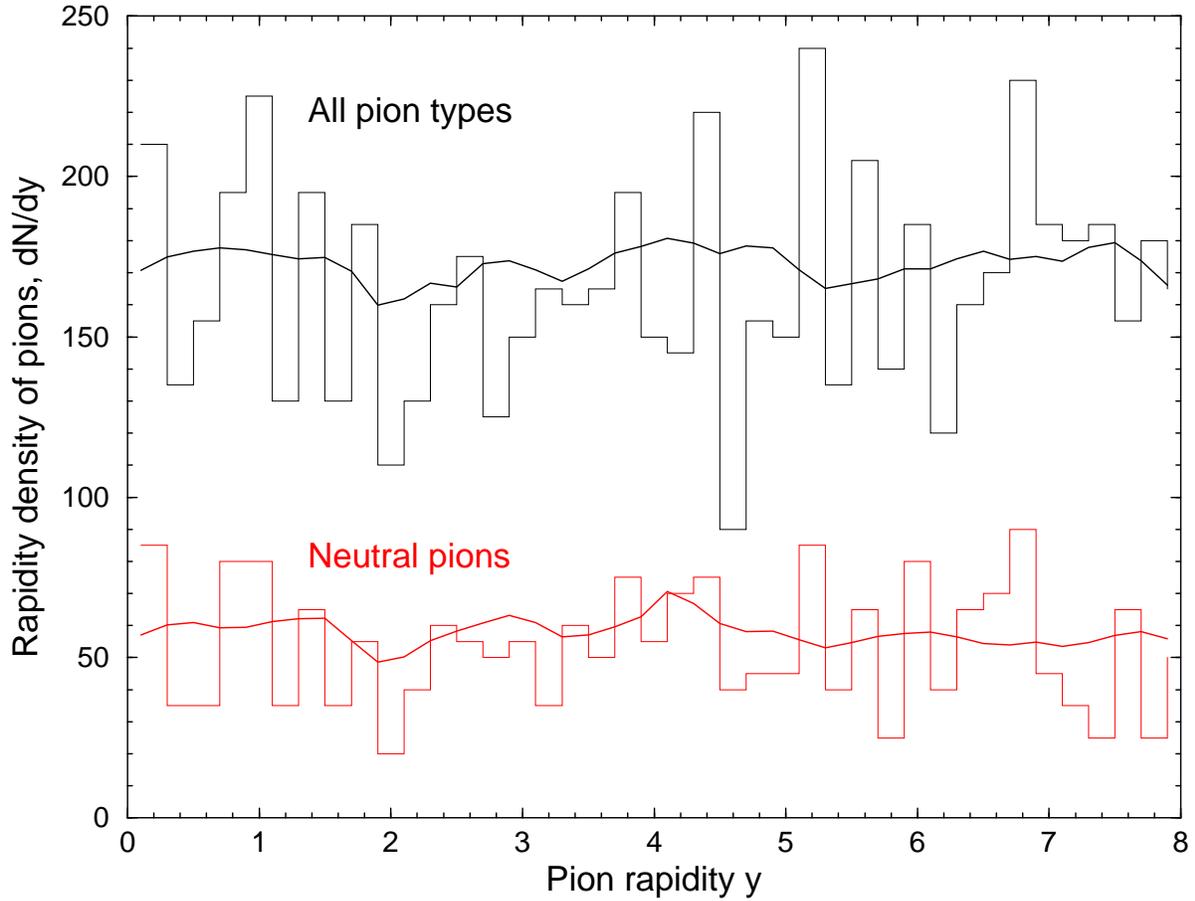}
\caption{Rapidity density.}
\label{f:dNdy}
{\small
The rapidity density of pions
for a single event prepared with $T_0=250~\MeV$,
as obtained by dividing the number in each slice
by its width $\Delta{\sf y}=0.2$.
The expected multiplicity based on the calculated asymptotic field
is indicated by the smooth solid curve,
while the histogram shows the actual number
as picked from the corresponding Poisson distributions.
The top curves include all three pion types,
whereas the bottom curves show the neutral pions only.
}
\end{figure}

As an illustration,
we show in Fig.\ \ref{f:dNdy} both the expected and the actual multiplicity
as obtained for a single event.
It follows from the above remarks
that the fluctuations in the calculated rapidity density
has two different sources.
One is the variation in the details of the resulting field configuration
from event to event and from one rapidity lump to another
(giving rise to the smooth curves in Fig.\ \ref{f:dNdy}).
The other is the subsequent Poisson sampling of the actual multiplicity
based on the smooth average given by the calculated final field
(leading to the fluctuating histogram in the figure).

In order to investigate whether the resulting multiplicity distributions
contain deviations from the relatively trivial Poisson statistics,
we have extracted the associated factorial moments.
Because of the enhancement of the soft pions seen in the transverse spectra,
we consider the pions below and above 200~MeV separately.
For multi-particle final states,
one may define the following factorial moments,
\begin{equation}
{\cal M}_m\ \equiv\ \prec N(N-1)\cdots(N-m+1)\succ\ .
\end{equation}
Here $N$ is the number of pions
emitted by a given rapidity lump of unit length
and the average is over all lumps
in a sample of events prepared in the same manner
(\ie\ same bulk temperature $T_0$ and rod radius $R_0$);
the contribution vanishes for a lump with a multiplicity smaller than
the particular order $m$.
For a Poisson multiplicity distribution characterized by 
the mean multiplicity $\bar{N}$,
the factorial moments are simply given by ${\cal M}_m=\bar{N}^m$.
Therfore we prefer to consider the reduced factorial moments,
${\cal M}_m/\bar{N}^m$,
which are then all unity for a Poisson multiplicity distribution.
Since $\bar N$ is not available experimentally,
we use instead the average of the observed multiplicities for each lump,
$\prec N\succ$.
This should provide a good approximation,
since the lumps are all macroscopically equivalent.

\begin{figure}
~\vspace{160mm}
\includegraphics{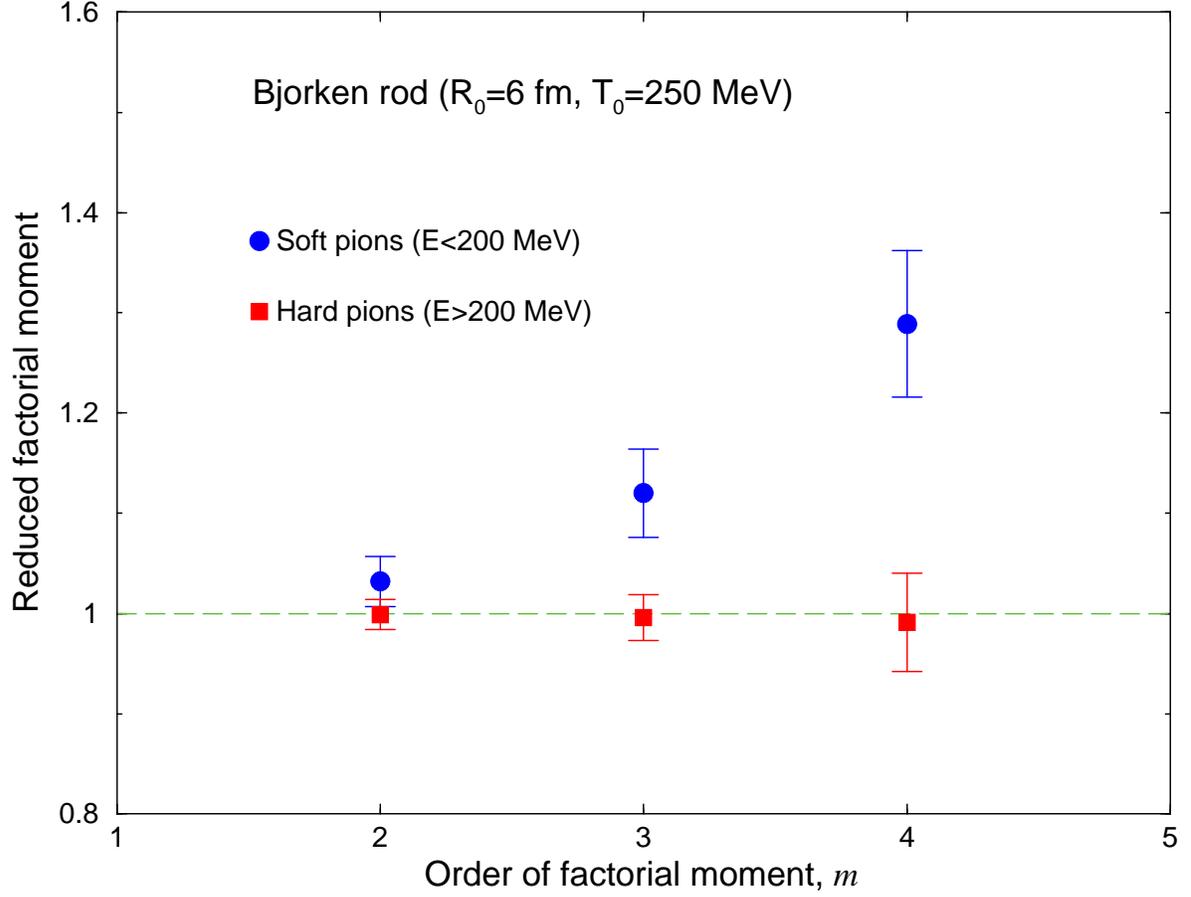}
\caption{Reduced factorial moments.}
\label{f:facmom}
{\small
The reduced factorial moments ${\cal M}_m/\bar{N}^m$
obtained for a sample of rods prepared with $R_0=6~\fm$ and $T_0=250~\MeV$
displayed as a function of the order $m$,
for either soft (dots) or hard (squares) pions.
The error bars arise from the sampling over all lumps and events
and the points have been slightly offset for clarity.
}
\end{figure}

Figure \ref{f:facmom} shows the relative factorial moments
obtained for a sample of Bjorken rods 
prepared with $T_0=250~\MeV$ and $R_0=6~\fm$.
(We consider from now on systems prepared with $T_0=250~\MeV$,
since the soft enhancement is largest near this temperature.)
While the hard pions appear to be perfectly consistent
with pure Poisson statistics,
the soft pions exhibit a significant non-poissonian behavior.
The character of the deviation of the soft factorial moments
suggests that the source occasionally emits anomalously many pions,
as one would expect if some modes are especially amplified.
This feature is consistent with the enhancement of the soft spectrum
and it can also be explored by other means of analyzing the data
for anomalous fluctuations.

\subsection{Azimuthal emission patterns}

When averaged over events,
the emission probability has overall cylindrical symmetry,
due to the symmetry of the initial ensemble of fields.
Similarly,
due to the boost invariance of the ensemble,
the averaged emission probability is independent of the rapidity.
And, thirdly, the isospin invariance of the ensemble
guarantees that all three types of pion behave similarly
after the event average has been carried out.
Though each individual lump retains these symmetries in a macroscopic sense,
texture exists at the microscopic level.
This important feature is manifested as fluctuations
in the underlying mean multiplicities $\bar{n}_\k^{(j)}({\sf y}_1,{\sf y}_2)$,
from one rapidity interval to the next and from event to event.

The fluctuations in the azimuthal pattern of the asymptotic pion field
can be elucidated by means of the following multipole moments,%
\footnote{The second-order multipole coefficient
gives a rough indication of the quadrupole moment of the emission pattern
and is related to the coefficient $v_2$ introduced in Ref.~\cite{v2}
for the discussion of elliptic flow.}
\begin{equation}
\tilde{\alpha}_M^{(j)}({\sf y}_1,{\sf y}_2)\ \equiv\ {1\over E_\perp}|
\sum_\k \bar{n}_\k^{(j)}({\sf y}_1,{\sf y}_2) E_k\ \rme^{iM\phi_\k}|\ .
\end{equation}
The weighting by the transverse kinetic energy $E_k\equiv m_k-m_\pi$
ensures that particles emitted with very low transverse momentum
have little effect,
as is reasonable since their particular azimuthal direction
is physically insignificant.
Because of the spectral enhancement of the soft pions discussed above,
we extract the multipole strength distribution
for soft and hard pions separately.
Finally,
$E_\perp\equiv\sum_\k\bar{n}_\k E_k$ denotes
the total mean transverse kinetic energy of the soft or hard pions
emitted by a given lump.
The division by $E_\perp$ ensures that $\tilde{\alpha}_0$ is always unity.
(It should perhaps be added that the rather large transverse dimensions
of the spatial lattice ensures that the corresponding resolution in the
wave number $\k$ is sufficiently fine to render the extracted multipole moments
coefficients numerically reliable for values of the multipolarity $M$ 
comfortably beyond the point when their values have become insignificant.)

In each individual event,
and for each of the three pion types separately,
the multipole moments are calculated for the pions
arising from each separate rapidity lump.
Because of the independence of different events,
the isospin symmetry,
and the longitudinal boost invariance, respectively,
all the individual lumps present equivalent sources
and hence we may perform a subsequent average,
yielding $\prec\tilde{\alpha}_M\succ$.

\begin{figure}
~\vspace{180mm}
\includegraphics{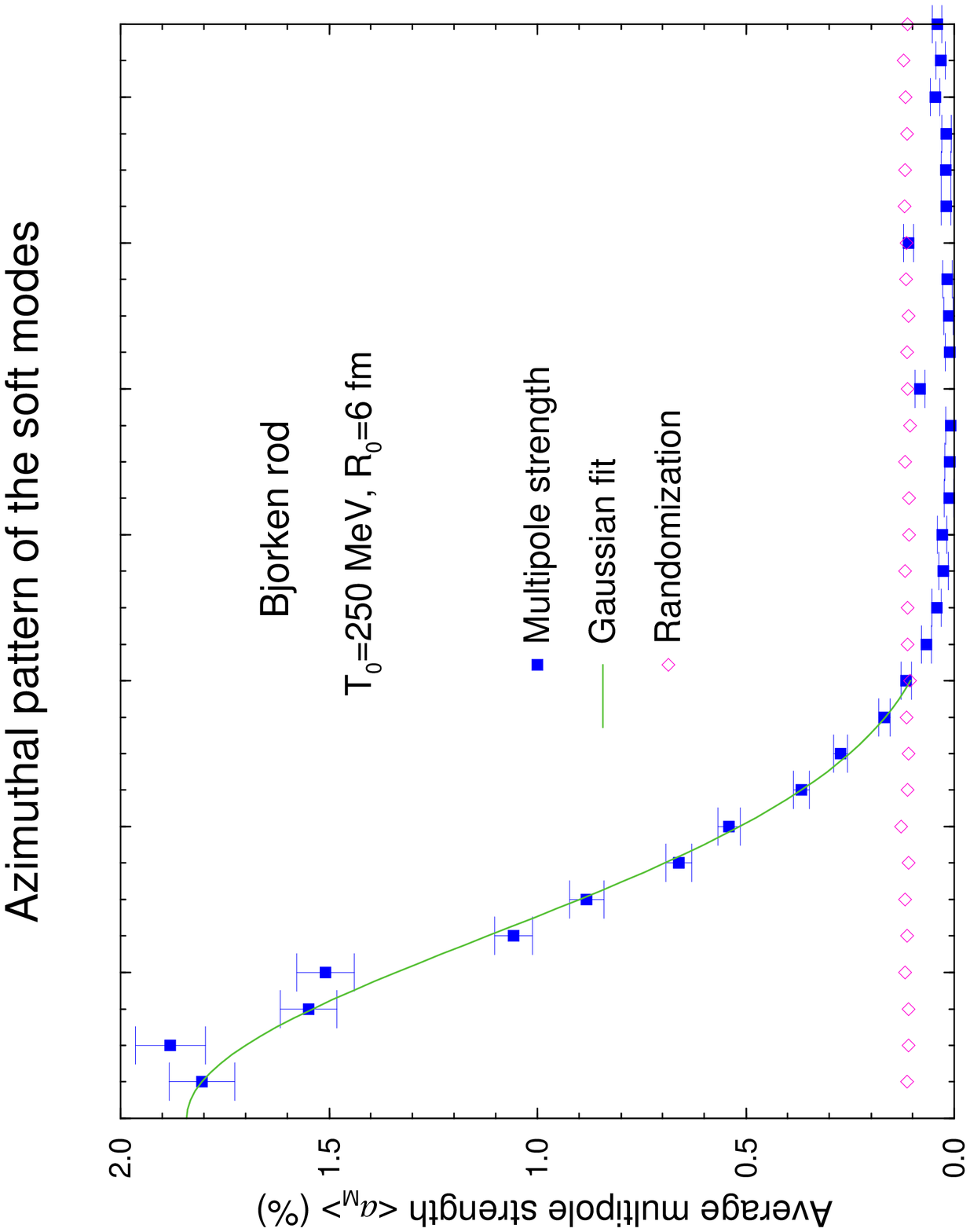}
\includegraphics{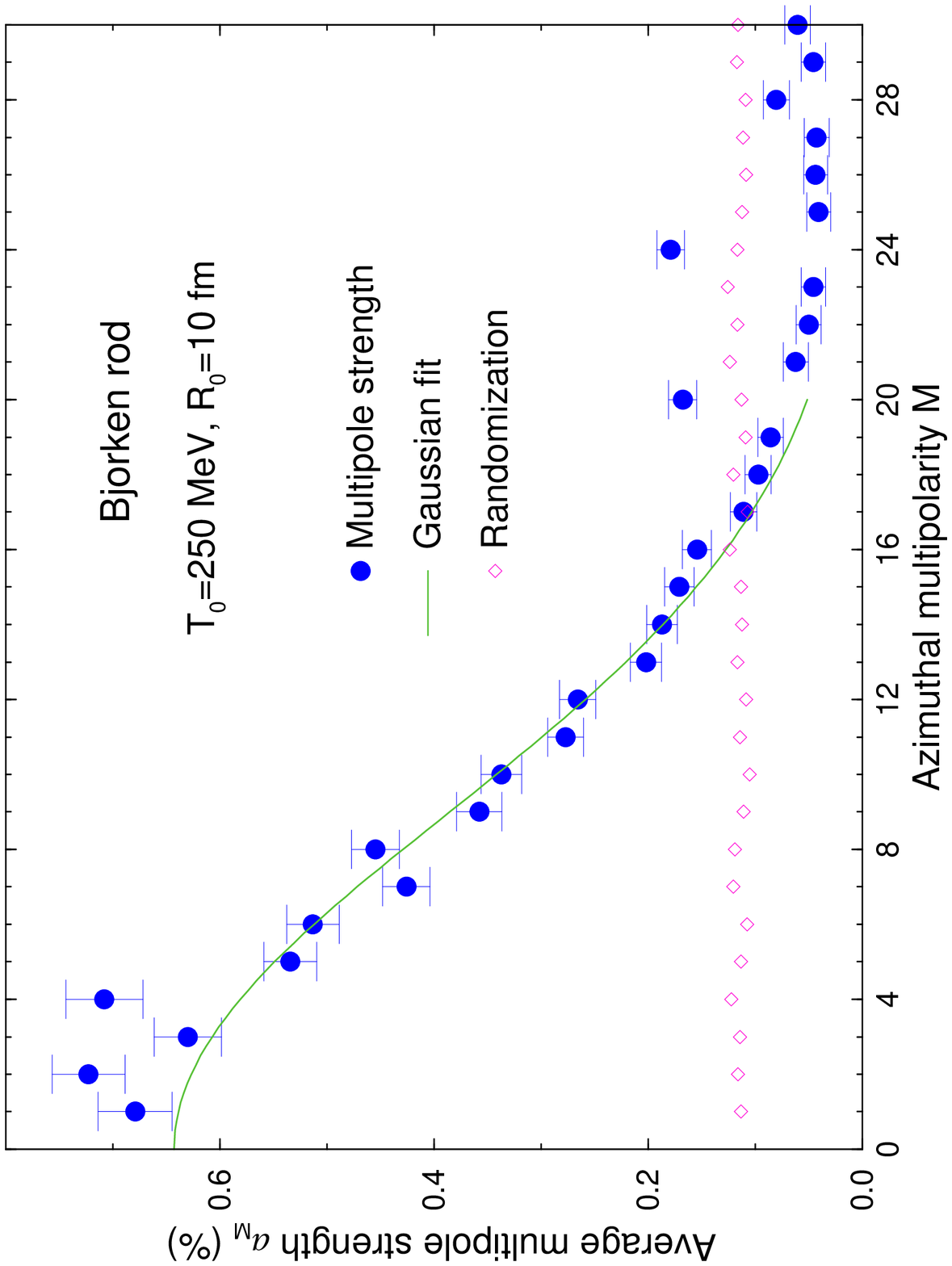}
\caption{Multipole strength for soft modes.}
\label{f:alphaL}
{\small
The azimuthal multipole strength distribution (in per cent) 
for the soft modes of the asymptotic pion field,
$\tilde{\alpha}$ (solid circles),
extracted for an ensemble of events
starting with an initial rod radius $R_0$ of either 6 or 10 fm
and an initial bulk temperature of $T_0=250~\MeV$.
The solid curve shows the corresponding gaussian fit. 
The open diamonds indicate the noise level
$\tilde{\alpha}^{\rm random}_M$,
obtained by randomizing the azimuthal direction
of each pion mode before evaluating the multipole moment.
}
\end{figure}

The resulting ensemble average azimuthal multipole strength function
is displayed in Fig.~\ref{f:alphaL}
for the soft modes of the asymptotic pion field.
Two different Bjorken rod scenarios are displayed,
both having the initial bulk temperature is $T_0=250~\MeV$,
but the initial radius $R_0$ being either 6~fm or 10~fm.
In order to get an idea of the significance of the extracted values,
we have calculated an auxiliary set of moments,
$\tilde{\alpha}_M^{\rm random}$,
obtained by turning each individual emission direction $\phi_\k$
by a random angle before evaluating the multipole moment.

The azimuthal coefficients for soft pions,
considered as a function of the multipolarity $M$,
start out well above the noise level,
then exhibit a gradual drop-off.
For high multipolarities,
the moments approach zero
because the high-frequency components of the field are suppressed by the
initial thermal weight.

The behavior of the multipole strength
provides some information on the domain structure of the source,
as we shall now discuss.
First we note that the attenuation of the strength
with increasing multipolarity $M$
can be reasonably well approximated by a simple gaussian fit.%
\footnote{In the numerical treatment,
the transverse pion momenta are given on a cartesian grid
which has a four-fold $C_4$ symmetry with respect to azimuthal rotations.
Consequently, those moments whose multipolarity $M$ is divisible by four
exhibit anomalous strength,
as is best seen for the high values of $M$.
We have therefore omitted those multipolarities from the fitting procedure.}
The fit makes it possible to extract the dispersion of the curve, $\Delta M$.
Employing the usual inverse relationship between conjugate variables,
we may then obtain the azimuthal correlation length,
$\Delta\phi=1/\Delta M$.
Thus, if the effective transverse size of the source is $R$,
the spatial domain size in the source is given by $\approx$$R\Delta\phi$.
For the two scenarios 
having the initial rod radius $R_0$ equal to 6 and 10 fm,
we extract $\Delta M$ as 5.03 and 8.84, 
leading to values of $\Delta\phi$ equal to
\mbox{11.4$\hspace{-0.82em}^\circ\hspace{0.4em}$} and 
\mbox{6.5$\hspace{-0.82em}^\circ\hspace{0.4em}$},
respectively.
Since $R_0\Delta\phi$ is thus approximately the same in the two scenarios,
the spatial correlation length at the effective emission time
appears to be independent of the initial radius of the rod.
This is as one would expect if that quantity reflects the conditions
in the source at the time of effective decoupling.

\begin{figure}
~\vspace{180mm}
\includegraphics{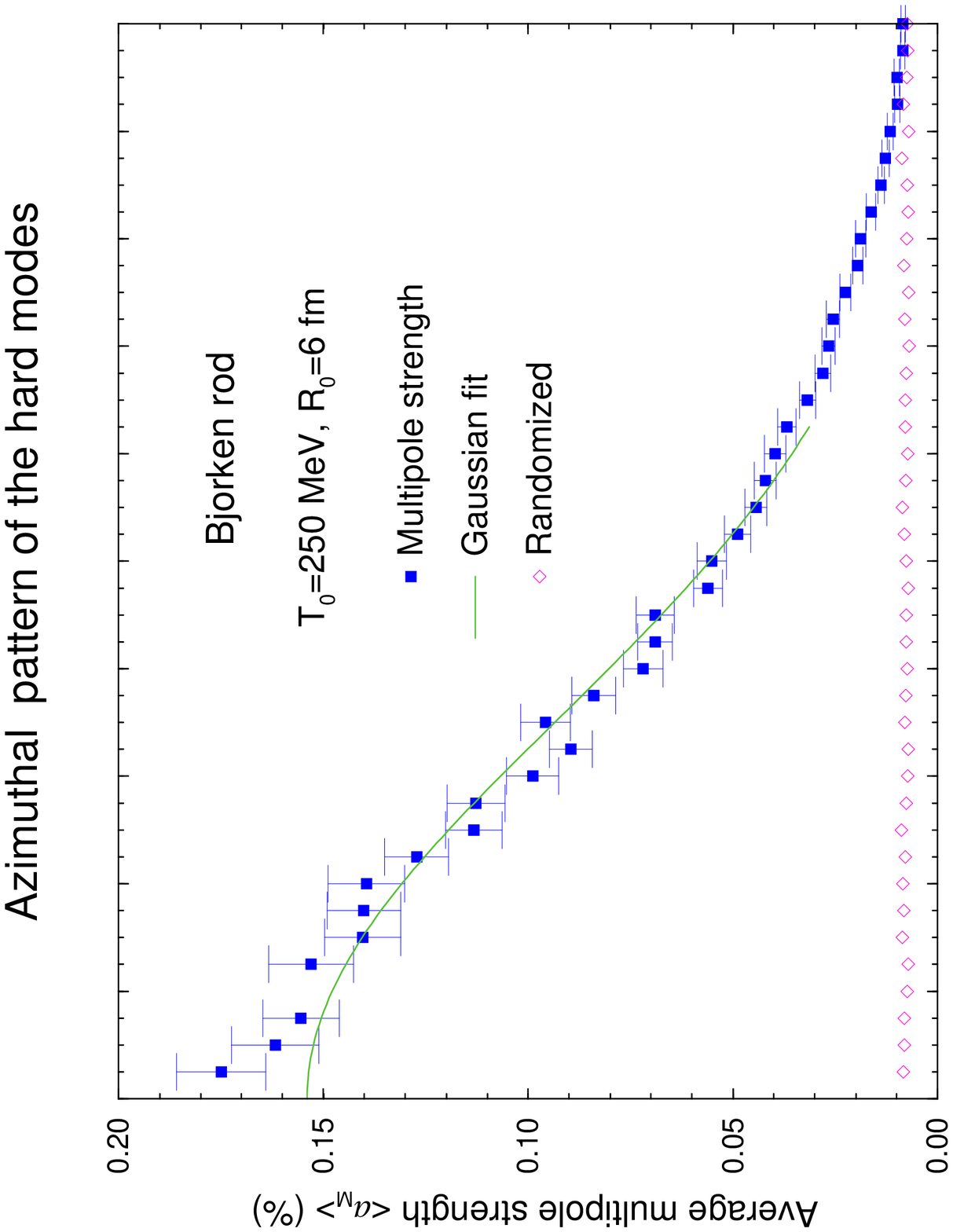}
\includegraphics{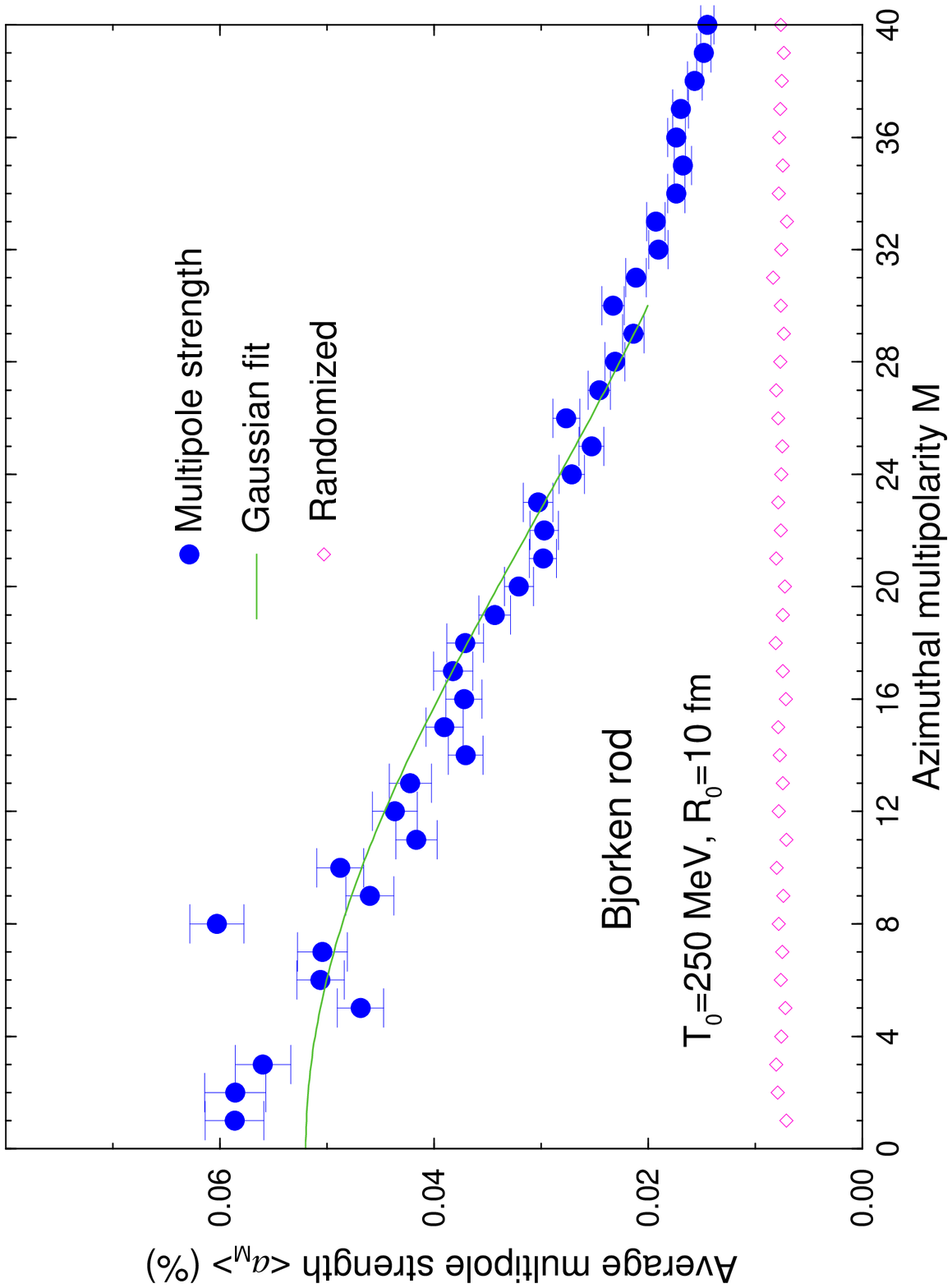}
\caption{Multipole strength for hard modes.}
\label{f:alphaH}
{\small
The azimuthal multipole strength (in per cent) 
of the hard modes of the asymptotic pion field, $\tilde{\alpha}$,
extracted for an ensemble of events
starting with an initial rod radius $R_0$ of either 6 or 10 fm
and an initial bulk temperature of $T_0=250~\MeV$,
together with the corresponding gaussian fit. 
The open diamonds indicate the noise level
$\tilde{\alpha}^{\rm random}_M$,
obtained by randomizing the azimuthal direction
of each pion mode before evaluating the multipole moment.
}
\end{figure}

The corresponding multipole coefficients extracted for the hard pions
are shown in Fig.~\ref{f:alphaH}.
They are initially an order of magnitude smaller than those for the soft pions
but they exhibit a significantly gentler fall-off with the multipolarity $M$,
suggesting a correspondingly narrower angular correlation.
This is consistent with the general expectation that harder pions
exhibit a smaller spatial correlation length.
As for the soft modes,
the wider rod yields smaller multipole coefficients.

The above discussion has been based on the pion {\em field},
rather than the resulting particles.
The field is a relatively smoothly varying function of position
and it governs the mean number of quanta in the various elementary modes.
The actual final state of particles is then determined stochastically,
as described in the preceding.
This feature tends to add a certain degree of noise to the azimuthal
pattern exhibited by the underlying field,
as we shall now illustrate.
When the analysis is made in terms of the emitted particles,
as is necessarily the case in actual experiments,
the definition of the multipole moments should be modified appropriately,
\begin{equation}
\alpha_M\ \equiv\ {1\over E_\perp}|\sum_n E_n\ \rme^{iM\phi_n}|\ ,\\
\end{equation}
where $n$ enumerates the $N$ particles emerging in the particular
kinematic domain under consideration
and $E_n\equiv m_n-m_\pi$ is the transverse kinetic energy
of an individual particle emitted in the azimuthal direction $\phi_n$.
Furthermore,
$E_\perp\equiv\sum_n E_n$,
is the total transverse energy of the $N$ particles,
so that $\alpha_0$ is unity.

\begin{figure}
~\vspace{180mm}
\includegraphics{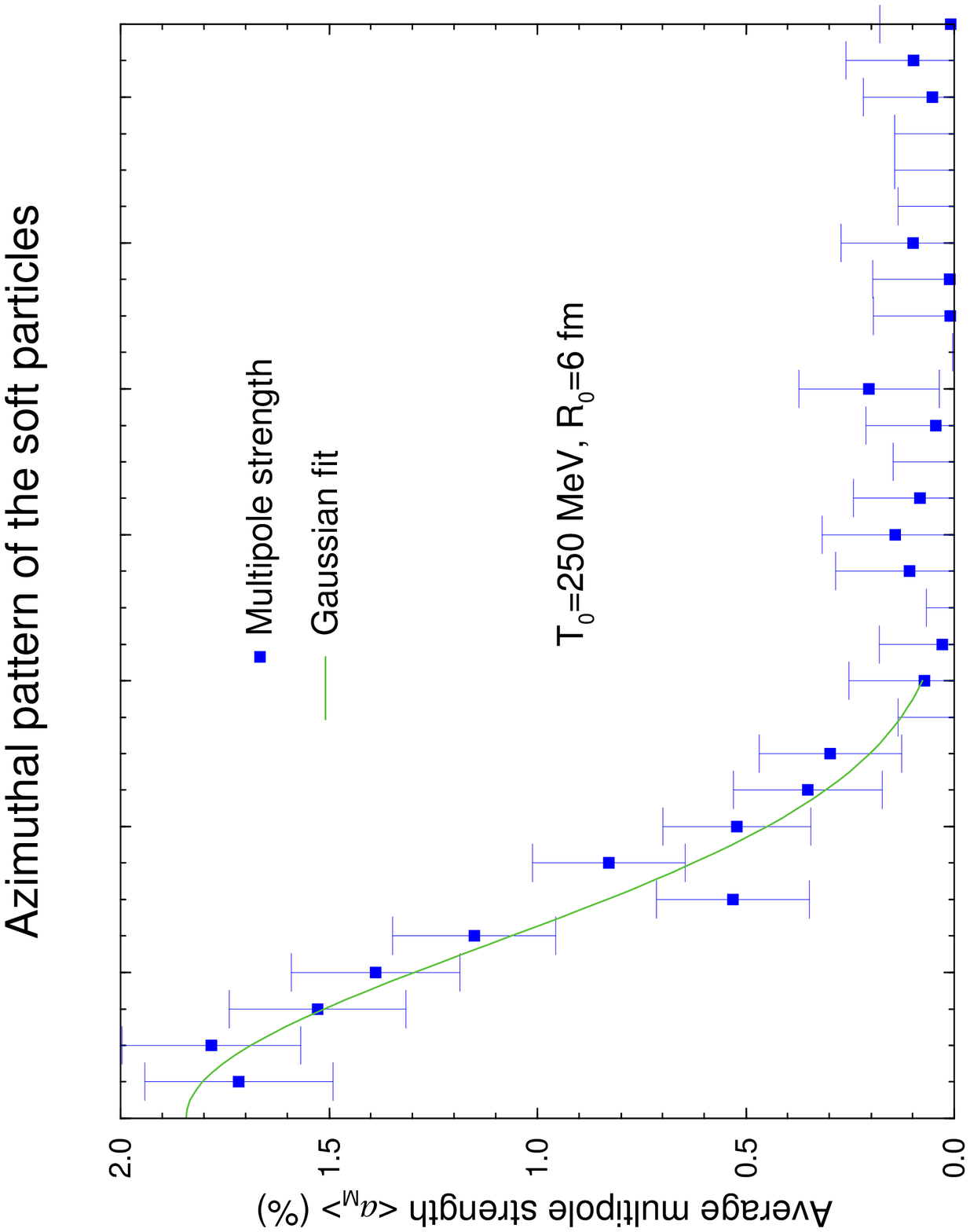}
\includegraphics{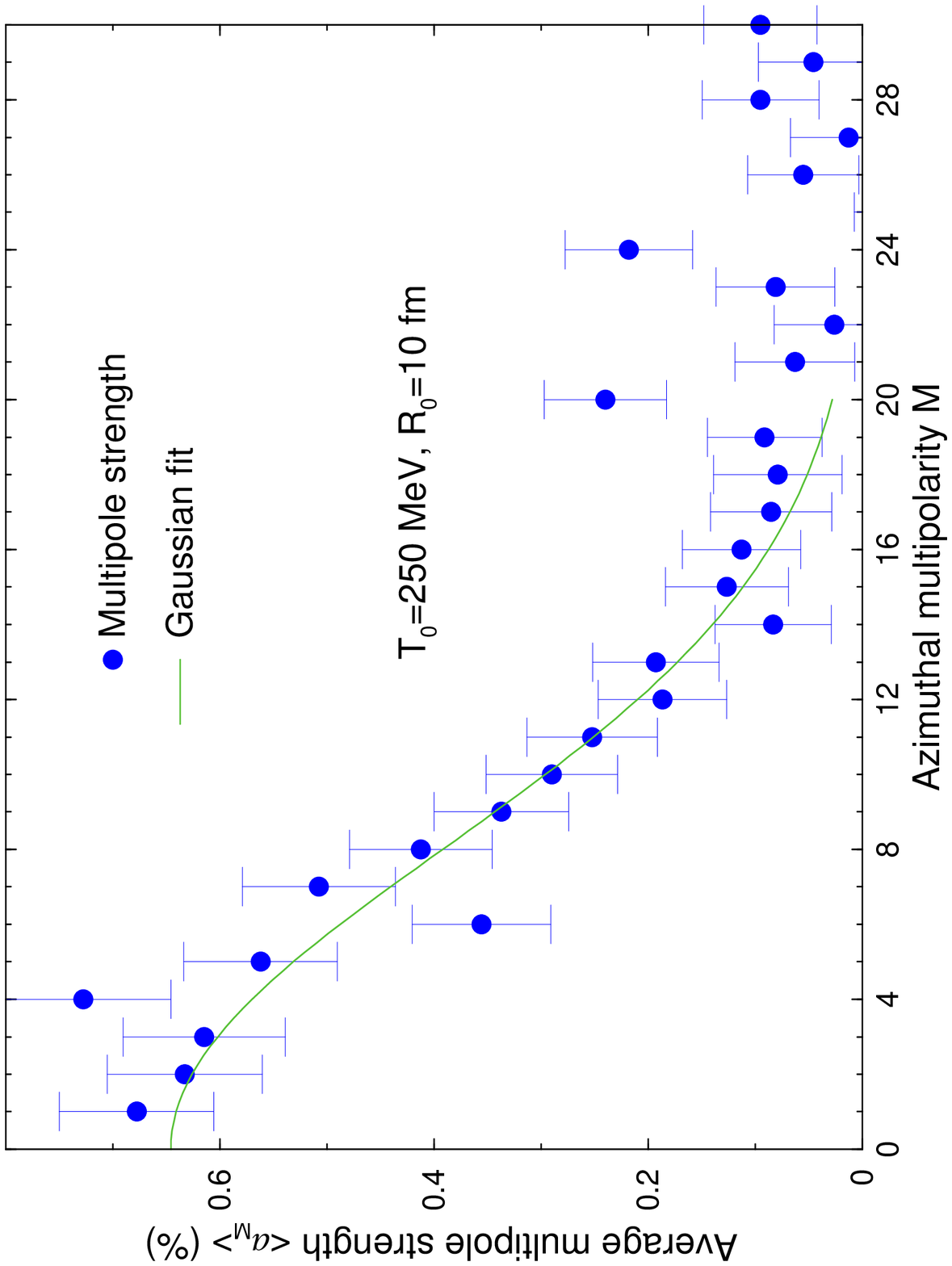}
\caption{Multipole strength for soft particles.}
\label{f:alphaP}
{\small
The azimuthal moments $\alpha_M$ (in per cent) extracted
for an ensemble of events starting with an initial rod radius $R_0=10~\fm$
and an initial bulk temperature of $T_0=250~\MeV$.
These moments have been extracted from the many-particle final states
obtained by sampling the actual particle numbers from
the underlying asymptotic fields used in Fig.~\ref{f:alphaL}.
The noise level averaged over the multipolarity $M$,
 $\langle\alpha^{\rm random}\rangle$,
has been subtracted from the displayed values
(they amount to 2.48 and 0.81 per cent for $R_0$ equal to 6 and 10 fm,
respectively)
and the solid curve shows the gaussian fit to the resulting difference. 
}
\end{figure}

Figure \ref{f:alphaP} shows the multipole strength coefficients $\alpha_M$
extracted for the case with $R_0=6~\fm$.
As was the case for the coefficients extracted from the underlying field,
the particle-based coefficients acquire significant values
for low values of the multipolarity $M$.
As the multipolarity is increased,
$\alpha_M$ approaches the statistical noise level, $\alpha^{\rm random}_M$,
obtained as above by randomizing the direction of each emitted pion.
This is what one would expect since the angular distribution is now a sum of
$\delta$-functions and thus contains all the high multipolarities equally.
To better bring out the features of the non-statistical behavior,
we subtract the average value of the noise,
$\langle\alpha^{\rm random}\rangle$,
before making the fit
(hence some of the resulting values are slightly negative for large $M$).

Once the average noise has been subtracted,
the multipole strength distribution based on the emitted particles
is quantitatively very similar to the one based on the underlying
asymptotic field.
This is true for both values of $R_0$ shown.
It thus appears possible to gain experimental access to the
correlation properties of the field itself,
even though the particles, on which the observations are necessarily based,
are obtained from the field by a stochastic process.

The above analysis suggest that an extraction of azimuthal moments
for observed soft pions may yield non-trivial information
on the domain structure in the system.
In this undertaking,
it would probably be useful to explore the effect of adjusting
the cuts in both rapidity and transverse energy,
so as to tune the extraction procedure to the specific character
of the sources produced.

\subsection{Neutral pion fraction}

The distribution of the neutral pion fraction $f\equiv N_{\pi^0}/N_{\rm tot}$
has received considerable attention as a DCC diagnostic
because an idealized fully isospin-polarized source would yield
an anomalously wide distribution, $P_0(f)=1/2\sqrt{f}$,
rather than a narrow peak around the average value $f=\third$.
Indeed,
experiments have been devoted to the search for anomalous behavior
of the neutral pion fraction \cite{MiniMax,WA98}.
However, such an undertaking is quite difficult,
partly because the signal is expected to be carried by only the soft pions
and partly because the extraction of $f$ requires the simultaneous
event-by-event measurement of both charged and neutral pions.
These practical problems not withstanding,
$P(f)$ remains a very instructive quantity in the DCC context
and we have therefore extracted it as well.

As above,
we consider soft and hard pions separately.
In each case, we calculate the neutral pion fraction $f$ for each rapidity lump
and construct its distribution $P(f)$ by binning the $f$ values obtained
from all lumps and events.%
\footnote{Since isospin symmetry guarantees that 
that the labeling of the three cartesian isospin axes may be permuted
without any physical effect,
the sampling error on $P(f)$ can be reduced 
by making the corresponding three entries into the histogram
from each individual source,
rather than merely a single entry \cite{JR:NPA}.
}
Since all extracted distributions $P(f)$ appear well-peaked 
around their common average of one third,
it suffices for our discussion to characterize a given distribution
by its variance $\sigma_f^2=\langle f^2\rangle-\langle f\rangle^2$.
It is elementary to show that the extreme idealized distribution $P_0(f)$
has the variance $\sigma_0^2=4/45$ and that
$N$ such sources of equal strength have the variance $\sigma_f^2=\sigma_0^2/N$
\cite{JR:NPA}.
More generally,
one may consider an assembly of $N$ different sources,
with individual distributions $P_i(f_i)$.
If the strength (\ie\ the total multiplicity) of a given source is $\nu_i$,
the neutral fraction for the entire assembly is given by
$f=\sum_i \nu_i f_i /\sum_i \nu_i$
and its distribution is therefore given by
\begin{equation}
P(f)\ =\ \left[ \prod_{i=1}^N \int df_i\ P_i(f_i)\right]\
\delta(f-\sum_{i=1}^N \nu_i f_i/\sum_{i=1}^N \nu_i)\ .
\end{equation}
It is easy to verify 
that the mean value of $f$ remains equal to one third.
Moreover, a slightly more elaborate calculation yields the
expression for the variance of the combined distribution,
\begin{equation}\label{sigmaf}
\sigma_f^2\ =\ { \sum_i \nu_i^2\over (\sum_i \nu_i)^2}\ \sigma_0^2\ .
\end{equation}
If the isospin polarizations of all the elementary modes $\k$ are totally
independent,
then the above formula (\ref{sigmaf}) applies and it is possible to predict
the resulting width $\sigma_f$ 
by substituting the mean multiplicities
$\bar{n}_\k=\sum_j \bar{n}_\k^{(j)}$ for the strengths $\nu_i$.
However,
in actuality the pion field changes smoothly from one mode to the next
and so the isospin directions of neighboring modes are somewhat correlated.
Consequently,
we expect the actual $\sigma_f$ to be significantly larger than the
lower bound provided by (\ref{sigmaf})
and indeed this turns out to be the case.

In the present discussion,
we calculate the neutral fraction both on the basis of the asymptotic fields,
which govern the mean multiplicities for each elementary mode,
and on the basis of the resulting particles,
which are picked stochastically from those mean multiplicities (see above).
Table \ref{t:Pf} lists the resulting widths $\sigma_f$
for Bjorken rods with radii $R_0$ equal to either 6 or 10 fm.
We notice that when the transverse size of the system is increased,
the distribution becomes narrower,
as expected since more domains can be accommodated across the source.
In addition to considering the usual lumps covering one unit of rapidity each,
we have also calculated $\sigma_f$ by separating each lump into its
five constituent slices.
Naturally, the associated decrease of the source size
causes $P(f)$ to broaden.
However, the the fact that the variance grows by less than a factor of five
is a reflection of the fact that the effective domain size extends beyond
a single slice, so that neighboring slices are not independent with regard to
their isospin alignment.
Finally,
we notice that the distributions based on the particles
(which are the observable quantities)
are generally significantly broader than those based on the underlying fields.

\begin{figure}
~\vspace{160mm}
\includegraphics{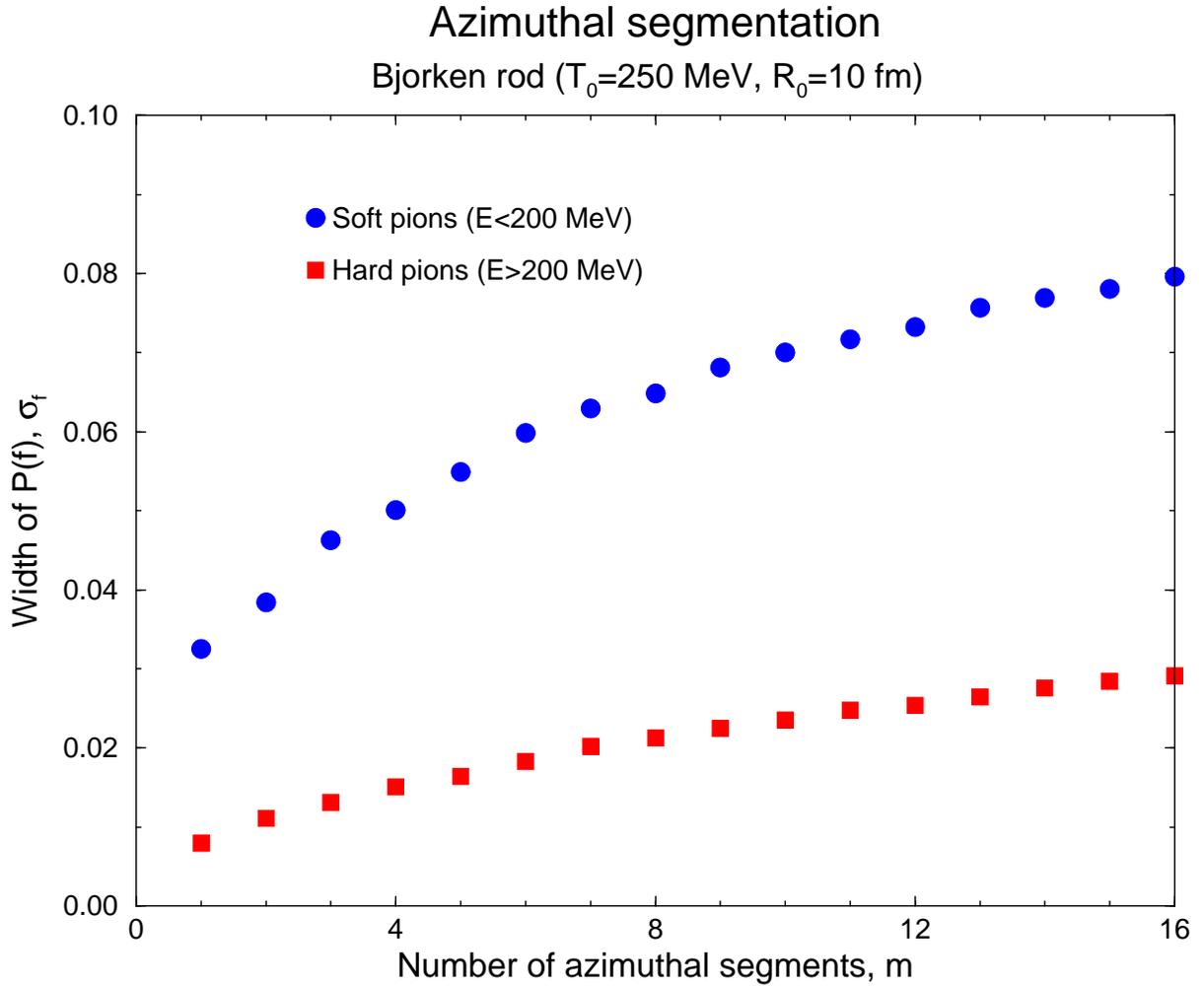}
\caption{Neutral pion fraction.}
{\small 
\label{f:AngDiv}
The dispersion $\sigma_f$ 
of the neutral pion fraction distribution $P(f)$
for sources that have been obtained by dividing each rapidity lump
into $m$ equally large azimuthal sections of width $2\pi/m$.
An ensemble of rods with a initial bulk temperature $T_0=250~\MeV$
and radius $R_0=10~\fm$.
The results for both soft and hard pions are shown.
These moments have been extracted from the many-particle final states
resulting from the underlying asymptotic fields used in Fig.~\ref{f:alphaL}.
}
\end{figure}

It is also instructive to study how $P(f)$ evolves
when the data is separated into azimuthal sections.
Thus, each individual lump is subdivided into $m$ equally large
azimuthal sections and $P(f)$ is calculated for each such source.
The result is illustrated in Fig.~\ref{f:AngDiv}
which shows the width of $P(f)$ as a function of the number of
angular sections $m$, for both soft and hard pions.
Generally $P(f)$ broadens as the width of the angular section is decreased,
as one would expect since the effective number of contributing domains
goes down.
However, $\sigma_f$ will not approach the value for a single idealized source,
as $m\to\infty$,
because the transverse extension of the source exceeds the correlation length
and so the source will never become fully aligned in isospace.
This feature is similar to what was obtained when idealized rods
were sliced ever finer in rapidity \cite{JR:PRD56}.
Moreover,
as we have seen above,
the soft pions lead to a significantly broader distribution
than the hard ones.

\section{Discussion}

In the present work,
we have employed the standard linear $\sigma$ model
to study systems prepared with a cylindrical geometry
and endowed with a longitudinal Bjorken scaling expansion.
Such a scenario is an idealized representation of what might be produced
in a central ultrarelativistic nuclear collision.
After their preparation with an interior temperature
in or above the chiral transition region,
the system is left free to evolve self-consistently,
until the combined longitudinal and transverse expansion
has brough it into the simple asymptotic regime of free dispersion
where a well-defined analysis can be made.
We have studied a number of observables related to the emerging pions,
for the purpose of exploring the persistence of signals reflecting
the preceding non-equilibrium relaxation of the chiral field.

Most important, and also most easily checked experimentally,\footnote{
It should be noted that the spectral profile is a one-particle observable
and thus easier to address experimentally than quantities requiring
true event-by-event analysis.}
is the emergence of a surplus of pions with transverse kinetic energies
below $200~\MeV$,
relative to the expectation based on a simple extrapolation 
of the harder part of the spectral profile,
which turns out to be quite well approximated by a Bose-Einstein form.
The excess of soft pions calculated in this manner
depends on the initial bulk temperature of the rod, $T_0$,
as well as on its radius, $R_0$,
and it is largest for $T_0=240-280~\MeV$.
In the optimal range of $T_0$,
the total relative excess of pions below $200~\MeV$
amounts to about $10\%$ and $15\%$ for $R_0$ equal to 6 and 10 fm,
respectively.
Since the transverse radius of the source is likely to be larger
than $10~\fm$ for central collisions at RHIC,
the present rod calculations are expected to underestimate the effect.
Thus,
if indeed the effect is present,
it ought to be readily visible in the data.

Since the additional soft pions arise from the oscillatory relaxation
of the chiral order parameter in the interior of the source,
they are expected to be non-statistical and partially isospin polarized.
In order to elucidate the degree to which remnants of these features
are identifiable in the final state,
we have performed various event-by-event analyses on the calculated results
based on samples of several tens of separate rod evolutions.

In one type of analysis, we have extracted the factorial moments
of the pion multiplicity distribution associated with a rapidity interval
of unit width.
We found that while the hard pions display a perfect Poisson multiplicity
distribution, the soft pions show a significant non-poissonian behavior,
with the reduced factorial moments growing approximately linearly with order.
A similar multiplicity analysis should be readily possible
with the multi-particle data collected by the STAR TPC, for example.

In another series of analyses,
we sought to elucidate the azimuthal structure of the final emission pattern
by extracting the corresponding multipole strength function.
These analyses were made both for the asymptotic field,
which is not directly detectable, and for the resulting particles.
The strength was found to decrease steadily with multipolarity
in an approximately gaussian manner,
thus making it possible to extract an angular correlation length.
These studies suggested that the spatial correlation length in the source
(the ``domain size'') is independent of the initial rod radius,
but significantly larger for the soft pions.
Moreover, these features were not affected by the stochastic transition
from field to particles,
thus suggesting that such angular analysis of the actual data
may be informative.

Finally,
we extracted the distribution of the neutral pion fraction
for the event samples considered,
since this quantity has played a central role in the discussion of
disoriented chiral condensates
(though it is practically difficult to measure).
Information about the correlation length,
and hence the domain structure,
was obtained by dividing each rapidity lump into azimuthal segments
before calculating the neutral fraction.
We found that the fraction distribution widens steadily
as the number of segments is increased and, importantly,
that the width associated with the soft pions is several times larger
than that for the hard pions.

Even though the present study is more refined than earlier treatments,
with regard to the dynamical scenario and the analysis of the outcome,
it is still highly idealized in a number of respects.
Especially important are the following limitations.
First,
at the most basic level,
the linear $\sigma$ model is an effective field theory
containing merely a few mesonic fields,
rather than the basic quark-gluon degrees of freedom.
This simplification may become especially important at the early stage
when the high excitation is expected to largely dissolve
the hadronic correlations in the interior of the system.
Second,
even at the effective level,
a full non-equilibrium real-time quantum-field treatment
is presently far beyond practicality.
Instead, we are resorting to a semi-classical treatment.
While perhaps quite adequate in many respects \cite{JR:LH98},
this treatment is expected to generally underestimate the enhancement
in the pion yield caused by the non-equilibrium dynamics \cite{JR:HIP}.
Third,
the commonly employed SU(2) version of the linear $\sigma$ model
entirely ignores strangeness.
However, 
this degree of freedom is expected to become significantly agitated
once the temperature exceeds the mass of the strange quark,
as is the case at the early stage of the evolution.
The extension of the model to SU(3)
has been found to have significant effects \cite{SU3}.
Fourth,
the initial state is obtained by sampling a (suitably modulated)
field configuration from a thermal ensemble.
The sampling procedure relies on the mean-field approximation
and thus the initial state contains
only a minimum of many-particle correlations.
This feature might become reflected in a corresponding paucity
of non-trivial correlations in the final state.
An analogous problem concerns
the transcription of the final classical field configuration
to a specific quantum state.
Since our prescription assigns a standard coherent state,
it is likely to provide an oversimplification
of the many-body correlations present.
(Indeed, recent studies suggest that the parametric amplification mechanism
behind the spectral enhancement in fact destroys the simple coherent form
of the state \cite{JR:HIP}.)
Fifth,
the present study addresses only part of the system.
A more realistic scenario would contain a halo of hadrons outside
the hot cylinder
and their interaction with the emerging pions might well be significant
(although the fact that the interesting signals
are carried by slow pions may give the surrounding material
time to disperse before having any significant effect).
In addition, of course, the cylindrical geometry is merely
an idealization of actual collision geometries occurring.
(In addition to these theoretical simplifications ,
our study ignores the deficiencies inherent in the detection system.)

Because of these various approximations and idealizations,
our calculations should not be considered as directly comparable
with experimental data. 
Rather,
our study serves to elucidate the importance of some of the complications
that are inevitably present in any real system
produced by high-energy collisions,
especially the rapid longitudinal expansion
and the presence of a surface region between the agitated interior
and the tranquil vacuum outside.
Generally speaking,
our results suggest that while one should not expect spectacular DCC signals
in the actual data, 
carefully designed quantitative analysis might
bring out any anomalous DCC-type behavior in the data.
In particular,
the observables considered in our analyses of the simulation results,
especially the transverse spectrum, the factorial moments,
and the multipole strength,
may be useful in this regard.




\begin{table}
\caption{The fitted BE temperatures.}
\label{t:TBE}
\begin{tabular}{|c|lllll|}
\phantom{space}$T_0~(\MeV)$\phantom{space}              
& 200  & 225 & 250 & 275 & 300 \\
\hline
$T_{\rm BE}~(\MeV)$     & 256  & 284 & 324 & 367 & 412 \\
\end{tabular}
~\\
{\small
The table shows the equivalent temperatures $T_{\rm BE}$
obtained by fitting a Bose-Einstein profile
to the calculated transverse pion spectra
in the kinetic-energy interval between 200 and 1000 MeV,
for Bjorken rods with an initial bulk temperature of $T_0$
and a radius of $R_0=6~\fm$.
Rods with $R_0=10~\fm$ lead to very similar values.
}
\end{table}

%
%
\begin{table}
\caption{Width of the neutral pion fraction distribution.}
\label{t:Pf}
\begin{tabular}{|l|ll|ll|}
Source type & 
\multicolumn{2}{l|}{\phantom{spacexxx}Fields} &
\multicolumn{2}{c|}{Particles\phantom{space}} \\
\hline
Source size& 
Lumps & Slices\phantom{space}& Lumps & Slices\phantom{space} \\
\hline\hline
Soft~ ($R_0$=6) &       5.34    & 6.51  & 11.74 & 26.98 \\
Hard ($R_0$=6)  &       1.29    & 1.82  & 8.15  & 19.54 \\
\hline
Soft~ ($R_0$=10)        &       3.25    & 4.14  & 6.66  & 14.16 \\
Hard ($R_0$=10)~        &       0.80    & 1.08  & 4.70  & 10.65 \\
\end{tabular}
~\\
{\small
The table shows the dispersion $\sigma_f$ of the neutral pion fraction
distribution $P(f)$ (given in per cent),
obtained for two ensembles of Bjorken rods
having an initial bulk temperature of $T_0=250~\MeV$
and a transverse radius $R_0$ equal to either 6 or 10 fm.
The neutral fraction has been calculated using either the
mean multiplicities (given by the field strength)
or the emerging particles (obtained stochastically based on the given mean).
The sources are taken to be either the lumps covering one unit of rapidity
or their  constituent slices which each cover $\Delta{\sf y}=0.2$.
Pions with transverse kinetic energies below (soft) and above (hard) 200 MeV
have been considered separately.
}
\end{table}


\newpage

\newpage

\newpage

\newpage

\newpage

\newpage

\newpage

\newpage

\newpage

\newpage

\newpage

\newpage

\newpage

\newpage

\newpage

\newpage

\newpage

\newpage

                        \end{document}